\documentclass{article} % For LaTeX2e
\usepackage{iclr2026_conference,times}

\usepackage{amsmath,amsfonts,bm}

\def\eqref#1{equation~\ref{#1}}
\def\1{\bm{1}}

\DeclareMathAlphabet{\mathsfit}{\encodingdefault}{\sfdefault}{m}{sl}
\SetMathAlphabet{\mathsfit}{bold}{\encodingdefault}{\sfdefault}{bx}{n}

\newcommand{\eg}{\textit{e.g.}}
\newcommand{\ie}{\textit{i.e.}}
\usepackage{amsmath,amsfonts}
\usepackage{microtype}
\usepackage{graphicx}
\usepackage{subcaption}
\usepackage{booktabs} % for professional tables
\usepackage{colortbl}
\usepackage{xcolor}
\usepackage{wrapfig}
\usepackage{pgfplots,pgfplotstable}
\pgfplotsset{compat=1.18}
\usepgfplotslibrary{polar}
\usepackage{multirow}
\usepackage{adjustbox}  % For table scaling

\usepackage{hyperref}
\usepackage{url}
\usepackage{bbding}
\usepackage{pifont}
\usepackage{wasysym}
\usepackage{amssymb}
\usepackage[capitalize,noabbrev]{cleveref}
\usepackage[most]{tcolorbox} % Add this in your preamble
\hypersetup{
    pdfborderstyle={/S/U/W 1}, % underline links instead of boxes
    linkbordercolor=red,       % color of internal links
    citebordercolor=green,     % color of links to bibliography
    filebordercolor=magenta,   % color of file links
   urlbordercolor=cyan        % color of external links
}

\definecolor{nvidiaGreen}{RGB}{118,185,0}
\definecolor{qwenPurple}{RGB}{116,81,207}
\definecolor{closedGray}{RGB}{90,90,110}

\usepackage{interval}
\usepackage{tabularx}
\usepackage{mwe}
\usepackage{xspace}
\usepackage{mathtools}  % for conditions in math
\usepackage{siunitx}    % for units

\usepackage{placeins}
\makeatletter
\renewcommand{\paragraph}{%
  \@startsection{paragraph}{4}%
  {\z@}{0.5em}{-5em}%
  {\normalfont\normalsize\bfseries}%
}

\makeatother
\usepackage{algorithm2e}
\usepackage{amsmath}
\usepackage{hyperref}
\usepackage{url}
\usepackage[framemethod=TikZ]{mdframed}
\usepackage{arydshln}
\usepackage{enumitem}

\usepackage{multirow}
\usepackage{colortbl}
\usepackage{xcolor}
\usepackage{amssymb}
\definecolor{xl}{HTML}{FAD9D5}
\definecolor{long}{HTML}{B9E0A5}
\definecolor{think}{HTML}{CCFFFF}

\title{
\raisebox{-0.2\height}{\includegraphics[width=0.06\textwidth]{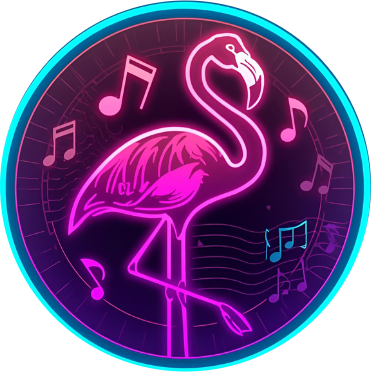}}%
\hspace{0.1em}
Music Flamingo: Scaling Music \\ Understanding in Audio Language Models
}

\author{Sreyan Ghosh$^{12*}$, Arushi Goel$^{1*}$, \textbf{Lasha Koroshinadze}$^{2**}$, \textbf{Sang-gil Lee}$^{1}$, \textbf{Zhifeng Kong}$^{1}$, \\ 
\textbf{Joao Felipe Santos}$^{1}$, \textbf{Ramani Duraiswami}$^{2}$, \textbf{Dinesh Manocha}$^{2}$, \textbf{Wei Ping}$^{1}$,\\ \textbf{Mohammad Shoeybi}$^{1}$, \textbf{Bryan Catanzaro}$^{1}$ \\ \\
NVIDIA, CA, USA$^{1}$, University of Maryland, College Park, USA$^{2}$ \\ \\
Correspondence: sreyang@umd.edu, arushig@nvidia.com}

\iclrfinalcopy % Uncomment for camera-ready version, but NOT for submission.

\begin{document}

\maketitle
\vspace{-9mm}
\begin{center}
    Project: \url{https://research.nvidia.com/labs/adlr/MF/}
\vspace{1mm}
\end{center}

\begin{abstract}
We introduce \textbf{Music Flamingo}, a novel large audio–language model, designed to advance music (including song) understanding in foundational audio models. While audio–language research has progressed rapidly, music remains challenging due to its dynamic, layered, and information-dense nature. Progress has been further limited by the difficulty of scaling \textit{open} audio understanding models, primarily because of the scarcity of high-quality music data and annotations. As a result, prior models are restricted to producing short, high-level captions, answering only surface-level questions, and showing limited generalization across diverse musical cultures. To address these challenges, we curate MF-Skills, a large-scale dataset labeled through a multi-stage pipeline that yields rich captions and question–answer pairs covering harmony, structure, timbre, lyrics, and cultural context. We fine-tune an enhanced Audio Flamingo 3 backbone on MF-Skills and further strengthen multiple skills relevant to music understanding. To improve the model's reasoning abilities, we introduce a post-training recipe: we first cold-start with MF-Think, a novel chain-of-thought dataset grounded in music theory, followed by GRPO-based reinforcement learning with custom rewards. Music Flamingo achieves state-of-the-art results across 10+ benchmarks for music understanding and reasoning, establishing itself as a generalist and musically intelligent audio–language model. Beyond strong empirical results, Music Flamingo sets a new standard for advanced music understanding by demonstrating how models can move from surface-level recognition toward layered, human-like perception of songs. We believe this work provides both a benchmark and a foundation for the community to build the next generation of models that engage with music as meaningfully as humans do.\footnote{$^*$Equally contributed and led the project. Names randomly ordered.$^{**}$ Significant technical contribution.}
\begin{center}
    \textit{``Without music, life would be a mistake.''} --- Friedrich Nietzsche
\end{center}

\end{abstract}

\section{Introduction} \label{sec:introduction}

\begin{figure}[!t]
    \centering
    \includegraphics[width=\linewidth]{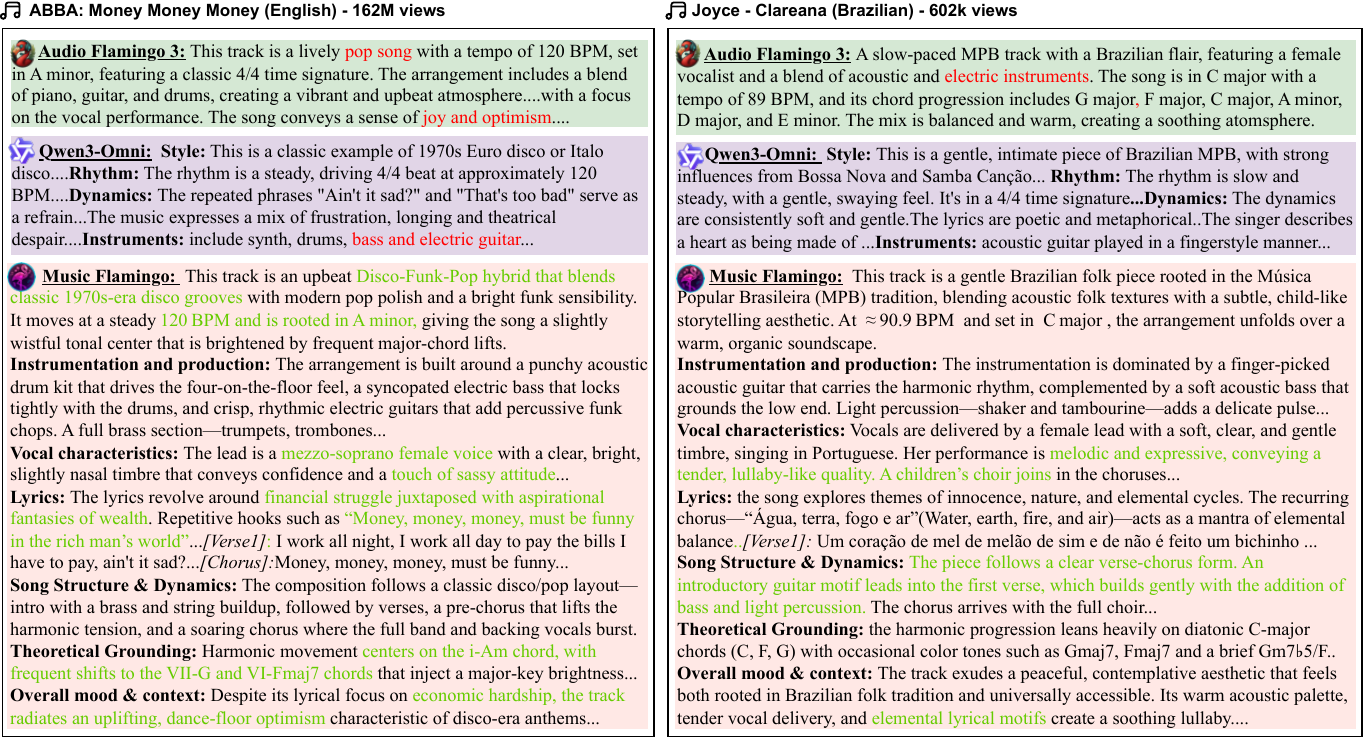}
    \vspace{-6mm}
    \caption{\small Comparison of captions for two diverse, full-length, in-the-wild songs by \textbf{Music Flamingo} and other frontier models. Prior models, such as AF3, tend to output short, surface-level descriptions (e.g., broad genre, tempo, or instrumentation), while Qwen3-Omni offers isolated observations without forming a coherent musical narrative. In contrast, Music Flamingo produces detailed, multi-layered captions that integrate theory-aware analysis with performance context. It links surface attributes (tempo, key, etc.) to mid-level structures (chord progressions, vocal phrasing, etc) and higher-level dimensions (lyrical meaning, emotional trajectory, etc.). This ability to connect one aspect of music to another results in richer, more holistic captions that resemble how trained musicians describe songs. Detailed expert analysis in Appendix~\ref{sec.user_study} and ~\ref{sec.user_study_cultures}.}
    \label{fig.comparison_figure}
    \vspace{-6mm}
\end{figure}
Audio -- including speech, environmental sounds, and music -- is central to human perception and interaction. It enables us to converse, perceive our surroundings, express emotions, interpret multimedia, and engage with cultural artifacts. Among these, music is particularly significant: the creation, sharing, discovery, and understanding of music are daily activities for billions worldwide. Recent progress in Audio–Language Models (ALMs) has extended language models into the auditory domain, enabling impressive advances in speech and sound understanding. Yet, music remains fundamentally distinct from other forms of audio. Core musical attributes such as key, tempo, harmony, instrumentation, and vocal styles are not present in non-musical audio and require specialized reasoning. Moreover, tasks adapted from speech and sound (\eg, captioning, transcription, retrieval) demand unique treatment when applied to music. To date, no model has achieved music understanding on par with the multi-modal breakthroughs seen in vision or speech. Improved music understanding would unlock richer applications in creation, recommendation, cross-cultural analysis, education, and interactive systems, enabling models to engage with music as deeply as humans do.

% Despite advances in scaling ALMs with larger models and datasets~\citepc{}, effective music and song understanding (hereafter, music and songs are refered to as just ``music'' in the remainder of the paper) remains an open challenge. As illustrated in Figure~\ref{}, current frontier LALMs, when asked to caption even widely recognizable tracks, tend to produce only short, generic descriptions. They often misidentify surface-level properties such as tempo and key, and fail to connect these to deeper semantic or affective meaning. We have also noticed cases where frontier models like Gemini~\citepc{} rely on text-derived knowledge rather than a genuine understanding of the music itself. We argue that \textit{this stems largely from the training data}: most music–text pairs originate from MusicCaps~\citepc{}, an early expert-annotated dataset, and the many synthetic datasets that followed inherit its stylistic and structural constraints due to models used to generate such synthetic data primarily being trained on MusicCaps. Finally, most datasets available online consist of short music-only snippets, with no vocals like actual songs, which hinders the learning of many more components described next.

Despite advances in scaling LALMs~\citep{goel2025audioflamingo3advancing,chu2024qwenaudio2,kimiteam2025kimiaudiotechnicalreport,tang2024salmonngenerichearingabilities,Bertin-Mahieux2011}, effective music understanding remains an open challenge (hereafter, we use ``music'' to refer broadly to both instrumental pieces and songs). Current frontier LALMs, when captioning even widely recognizable tracks, often produce short and generic descriptions, misidentify surface-level attributes such as tempo or key, and sometimes rely on text-derived knowledge rather than genuine auditory analysis~\citep{comanici2025gemini25pushingfrontier}. We argue this stems largely from data: most available music–caption pairs originate from early datasets like MusicCaps~\citep{agostinelli2023musiclmgeneratingmusictext}, and subsequent datasets inherit its stylistic limitations of short, surface-level summaries, limitations of short, surface-level summaries that omit bar/time localization, harmonic and formal structure, vocal/lyric grounding, and cultural context, and often with a narrow focus on instrumental-only snippets. This prevents models from learning the layered nature of music, spanning surface attributes (tempo, key, timbre), mid-level structures (chord progressions, rhythm, phrasing), and higher-level dimensions (lyrics, emotional arcs, cultural context). Architecturally, training practices for most music LLMs or captioners still constrain holistic learning -- for example, the use of encoders (like CLAP~\citep{elizalde2022clap}) that do not capture spoken content or low-level features like pitch in their representations (see our study in Appendix~\ref{sec.linear_probing}), thereby constraining learning of vocal timbre, lyrical alignment, and expressive nuances in songs. We contend that even a task as basic as music captioning, when re-imagined beyond surface-level summaries, is inherently explorative and compositional: a musically informed description requires reasoning through multiple layers of structure and meaning, and admits not one single answer but a spectrum of valid interpretations shaped by theory, perception, and artistry.

{\noindent \textbf{Main Contributions.}} In this paper, we introduce \textbf{Music Flamingo}, a new and \textit{open-source} large audio–language model specifically designed to advance music understanding. Unlike speech or environmental sounds, music is inherently \textit{layered, expressive, and structured}, combining surface-level acoustic attributes (tempo, key, timbre) with mid-level organization (harmony, form, rhythm) and higher-level dimensions (lyrics, style, affect, cultural context). Capturing this multi-faceted nature of music requires models that can move beyond surface-level recognition toward reasoning and interpretation more akin to a trained musician.  

To build Music Flamingo, we re-imagine the scope of music understanding and recast conventional tasks, such as music captioning and question answering, into comprehensive formulations that demand deliberate, step-by-step reasoning (Fig.~\ref{fig.mf_skills}). To support this reframing, we introduce new strategies for both data curation and model training. First, we present \textbf{MF-Skills}, a dataset with 4M+ high-quality samples for training music-understanding models. Unlike prior corpora dominated by short, instrumental snippets, MF-Skills scales to long, multicultural full-length songs with vocals drawn from diverse sources. We propose a multi-step labeling pipeline that yields detailed, multi-aspect, \emph{layered} captions -- capturing harmony, structure, timbre, lyrics, and cultural context -- designed to elicit musician-level reasoning. Beyond captions, MF-Skills includes carefully curated question–answer pairs that move past simple instrument identification toward tasks requiring temporal understanding, harmonic analysis, lyrical grounding, and other skills. On the modeling side, we first identify core limitations in Audio Flamingo 3 and continue-pre-training it to build a stronger backbone by fine-tuning it on multilingual, multi-speaker ASR and extended audio reasoning datasets before specializing it for music. Next, we propose a post-training stage specifically designed to enhance reasoning. For this stage, we further introduce \textbf{MF-Think}, a dataset of 300K chain-of-thought examples grounded in music theory, which we use for cold-start reasoning training. Finally, we apply GRPO-based reinforcement learning with custom rewards, enabling explicit step-by-step musical reasoning. In summary, our contributions are:

\begin{itemize}[leftmargin=*, labelsep=1em, itemsep=0pt, topsep=0pt]
\vspace{-2mm}
\setlength\parskip{0em}
    \item We propose \textbf{Music Flamingo}, a new LALM for advancing music understanding. We re-imagine conventional music tasks (\eg, captioning, QA) as reasoning-centric formulations and introduce novel training strategies tailored to these tasks.  
    \item To support training, we release \textbf{MF-Skills} and \textbf{MF-Think}, two large-scale datasets containing music–caption and music–QA pairs designed to promote deliberate reasoning. Unlike prior datasets limited to short instrumental clips, ours include full-length, multi-cultural songs with detailed, multi-aspect annotations.  
    \item Music Flamingo achieves state-of-the-art results on $12$ music understanding and reasoning benchmarks. Beyond academic benchmarks, expert evaluations show its outputs are more accurate and preferred by trained musicians than existing models.  
    \item To promote research in this area, we will release code, training recipes, and our new datasets under an appropriate research-only license.  
\end{itemize}
% \vspace{-1mm}

\section{Related Work}
\label{sec:related_work}
% \vspace{-2mm}
\noindent \textbf{Multimodal audio–language modeling.} The rapid progress of LLMs has accelerated the development of multimodal LLMs (MLLMs) capable of understanding and reasoning across diverse modalities, including audio. Within this space, ALMs focus specifically on reasoning over auditory inputs such as speech, sounds, and music. Architecturally, ALMs generally follow two paradigms:
(i) \textit{Encoder-only ALMs}, which learn a joint embedding space for audio and text, enabling tasks like cross-modal retrieval. Representative models include CLAP~\citep{elizalde2022clap}, Wav2CLIP~\citep{wu2021wav2clip}, and AudioCLIP~\citep{guzhov2021audioclip}.
(ii) \textit{Encoder–decoder ALMs} (often called Large Audio–Language Models, LALMs), which augment decoder-only LLMs with audio encoders. Notable examples include LTU~\citep{gong2023ltu}, LTU-AS~\citep{gong2023ltu-as}, SALMONN~\citep{tang2024salmonngenerichearingabilities}, Pengi~\citep{deshmukh2023pengi}, Audio Flamingo~\citep{kong2024audioflamingo}, Audio Flamingo 2~\citep{kong2025audioflamingo2}, Audio Flamingo 3~\citep{goel2025audioflamingo3advancing}, AudioGPT~\citep{huang2023audiogpt}, GAMA~\citep{ghosh2024gama}, Qwen-Audio~\citep{chu2023qwenaudio}, and Qwen2-Audio~\citep{chu2024qwenaudio2}. There has also been a surge of LALMs that specifically focus on music, including Mu-LLaMA~\citep{liu2024music}, MusiLingo~\citep{deng2024musilingobridgingmusictext}, M2UGen~\citep{liu2024m2ugenmultimodalmusicunderstanding}, SALMONN~\citep{tang2024salmonngenerichearingabilities}, and LLARK~\citep{gardner2024llarkmultimodalinstructionfollowinglanguage}. These LALMs have substantially advanced core audio understanding tasks such as automatic speech recognition (ASR)~\citep{radford2022whisper}, audio captioning~\citep{kim2019audiocaps}, and acoustic scene classification~\citep{chen2022beats}. More importantly, they have enabled \textit{open-ended audio question answering}, which requires both complex auditory reasoning and external world knowledge. While music has often been included as a modality within these models, it has rarely been a central focus.

Scaling music understanding within ALMs has proven particularly difficult. For instance, while the Audio Flamingo series has expanded its training data substantially from version 1 to 3, the music component of training data has increased by only $\approx$10\%, compared to much larger growth in speech and environmental sounds. Similarly, models such as Kimi~\citep{kimiteam2025kimiaudiotechnicalreport} and Step Audio~\citep{huang2025step} (where training data disclosures exist) show comparable imbalances. Finally, models like LLARK~\citep{gardner2024llarkmultimodalinstructionfollowinglanguage} and MU-LLaMA~\citep{liu2024music} curate music captions and question-answer pairs from existing open-source datasets, which lack diversity and skills. This is due to several challenges: the difficulty of collecting high-quality and culturally diverse music audio~\citep{kumar2025mmauprochallengingcomprehensivebenchmark}, curating reliable annotations~\citep{christodoulou2024multimodal}, and the reliance of most works on private, proprietary datasets~\citep{agnew2024soundcheckauditingaudio}. Large labs often construct in-house collections of lyrics and metadata by scraping online lyric repositories, translations, and song databases~\citep{ahmed2025sleepingdisco9mlargescalepretraining}. Models such as Jukebox and Neural Melody Reconstruction exemplify this paradigm. Finally, as noted earlier, most publicly available datasets emphasize short instrumental clips, with very limited coverage of full-length songs containing vocals, hindering a comprehensive understanding of music~\citep{kumar2025mmauprochallengingcomprehensivebenchmark}. 

\noindent \textbf{Music information retrieval and captioning.} Beyond LALMs, music understanding has a long history in Music Information Retrieval (MIR), encompassing retrieval, classification, and captioning. Foundational tasks such as key detection~\citep{chai2005detection}, chord recognition~\citep{sheh2003chord}, and tempo estimation~\citep{scheirer1998tempo} have been extensively studied, largely in instrumental music. Lyrics transcription has also been explored, posing a more difficult challenge than ASR due to overlapping vocals, diverse singing styles, and background instrumentation~\citep{mesaros2010automatic}. As discussed earlier, music captioning has been studied in analogy to audio event captioning, but typically produces short, high-level semantic descriptions rather than layered, theory-aware accounts. Importantly, improved captioning not only benefits downstream music understanding but also supports the training of generative music models by providing high-quality text supervision for in-the-wild audio~\citep{agostinelli2023musiclmgeneratingmusictext}. This connection has recently been emphasized in both standalone music modeling efforts and in broader video generation systems~\citep{chen2025mv}. 
% \vspace{-2mm}

% \vspace{-2mm}
\section{Methodology}
\label{sec:approach}
% \vspace{-2mm}

\begin{figure}[!t]
    \centering
    \includegraphics[width=\linewidth]{
    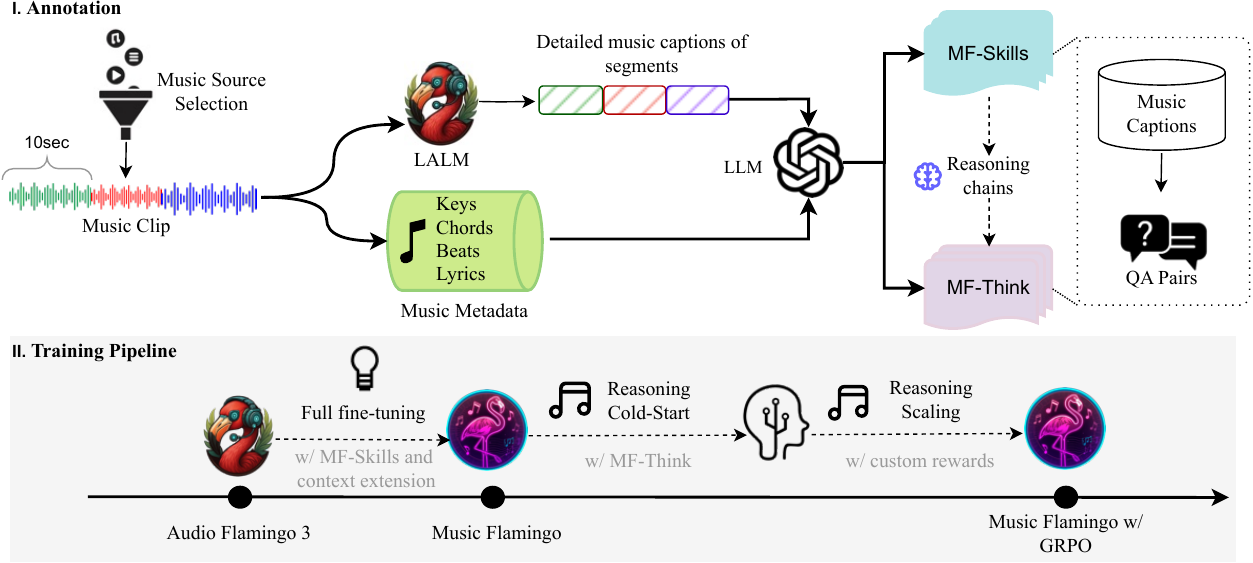}
    \vspace{-2mm}
    \caption{\small \textbf{I. Annotation pipeline} for constructing our proposed datasets from diverse music clips. \textbf{II. Training pipeline} of Music Flamingo: we begin by improving Audio Flamingo 3, then perform full fine-tuning on music datasets to build the Music Flamingo foundation model. Finally, the model undergoes reasoning cold-start training followed by GRPO fine-tuning to enable step-by-step reasoning.}
    \label{fig.mf_method}
    \vspace{-4mm}
\end{figure}

To build Music Flamingo, we first curate high-quality songs, followed by labeling them and finally fine-tuning the model on the curated data. Music Flamingo is a specialized music understanding model built by fine-tuning a version of Audio Flamingo 3, specifically with high-quality data to close the gap on skills and tasks crucial for music understanding. Finally, the model is fine-tuned using reinforcement learning to enable step-by-step music reasoning.  
\vspace{-2mm}
\subsection{Improved Audio Flamingo 3 Baseline} 
\label{sec.improved_af3}
\vspace{-2mm}
\noindent \textbf{Data.} We first strengthen Audio Flamingo 3 to serve as the backbone for Music Flamingo. Unlike instrumental-only music, songs contain vocals that contribute not only lyrics but also timbre, style, and expressive variation. Capturing these elements requires stronger spoken language understanding than prior baselines. Thus, in addition to the data used for AF3 training, we add the following to the mix: 1) Across all fine-tuning stages (1–3), we incorporate large-scale multilingual ASR data (sources Emilia dataset~\citep{he2024emiliaextensivemultilingualdiverse}, CoVoST~\citep{wang2020covost2massivelymultilingual}, MUST~\citep{qin2025mustdatasetunifiedframework}, Amazon-SIFT~\citep{pandey2025sift50mlargescalemultilingualdataset}; details in Appendix~\ref{sec:appendix}) to better capture global vocal diversity,  2) In stage 3, we add multi-talker ASR data, including  CHIME~\citep{watanabe2020chime6challengetacklingmultispeakerspeech,cornell2023chime7dasrchallengedistant}, Switchboard~\citep{godfrey1992switchboard} and ALI meeting~\citep{yu2022m2meticassp2022multichannel}, enabling the model to parse turn-taking and overlapping voices, which is critical for understanding duets and ensemble singing and, 3) We expand the data mix with speech-centric skills, including phoneme recognition and lyrics transcription, improving alignment between vocal content and musical context.  

\noindent \textbf{Training Pipeline.} We adopt the training paradigm introduced in Audio Flamingo 3~\citep{goel2025audioflamingo3advancing} to fine-tune the model on the diverse set of speech data described above. The resulting fine-tuned model then serves as the foundation for developing the music-focused foundational model. 
\begin{figure}[!t]
    \centering
    \includegraphics[width=\linewidth]{
    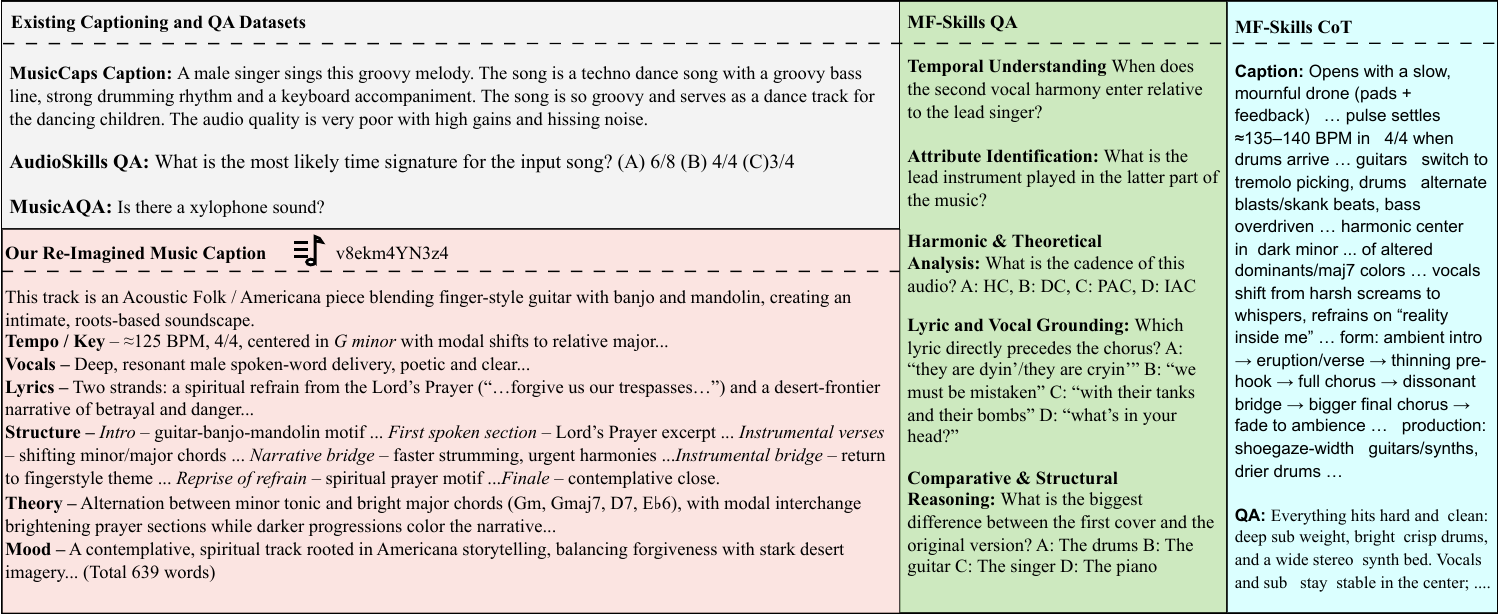}
    \vspace{-4mm}
    \caption{\small Examples from \colorbox{xl}{MF-Skills Caption}, \colorbox{long}{MF-Skills QA}, and \colorbox{think}{MF-Think}. We emphasize that our re-imagined captions are denser, more informative, and designed to require deliberate reasoning to generate. Additional examples are provided in Appendix~\ref{subsec:examples_mfthink}.}
    \label{fig.mf_skills}
    \vspace{-4mm}
\end{figure}
\vspace{-4mm}
\subsection{Building Foundational Music Understanding} 
\label{sec:bulding_music}
\vspace{-1mm}

\noindent \textbf{MF-Skills.} Prior captioning datasets mostly provide surface-level summaries, while existing QA \begin{wrapfigure}{r}{0.5\textwidth}
    \vspace{-0.6em} % optional: nudge up to align with paragraph
    \centering
    \includegraphics[
        clip,
        width=\linewidth
    ]{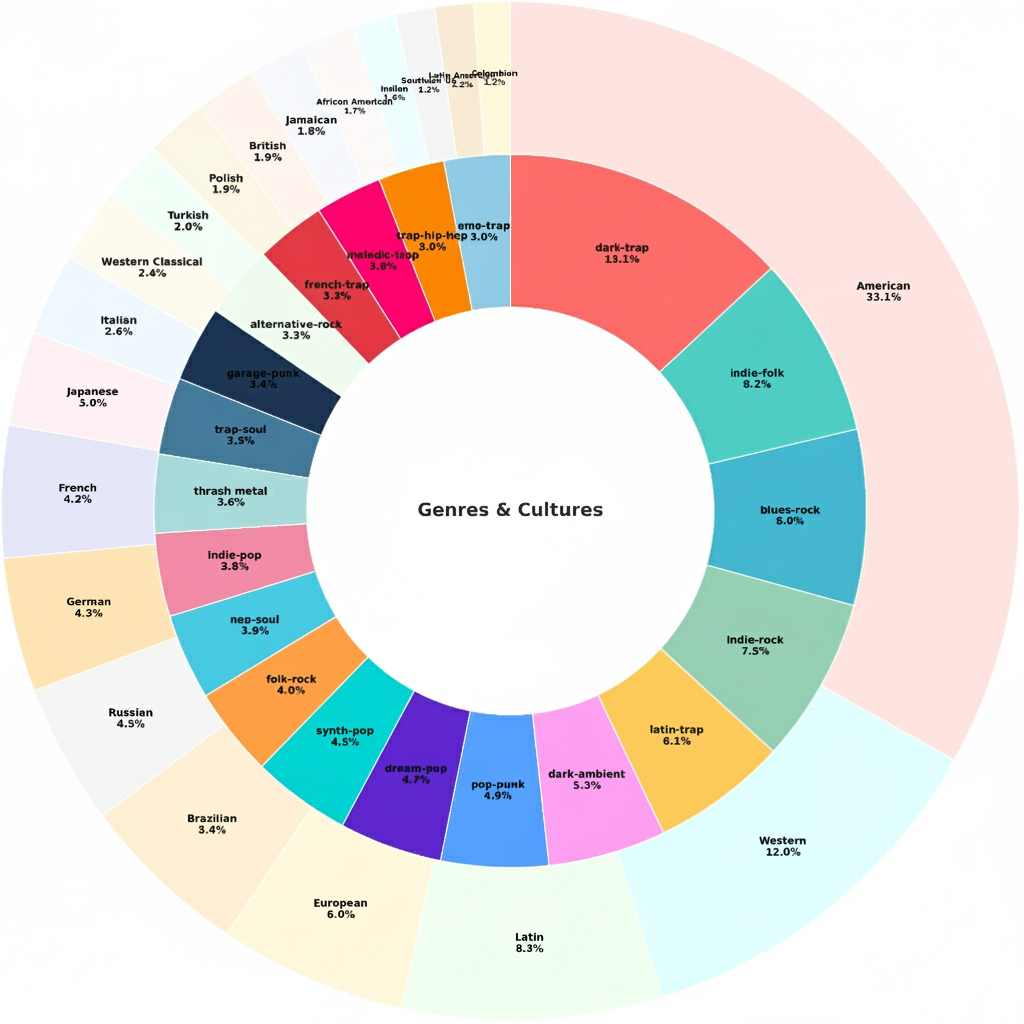}
    \vspace{-1em}
    \caption{\small Genres (inner circle) \& Cultures (outer circle) distribution of songs.}
    \label{fig:donut}
    \vspace{-2em}
\end{wrapfigure}datasets are dominated by simple classification tasks (\eg, instrument or tempo detection). Even large-scale skill-focused datasets such as AudioSkills focus primarily on sounds and speech for their diverse skill-specific QAs, with music data reduced to basic information extraction. To mitigate this gap, we design \textbf{MF-Skills} and capture the layered nature of music to train models for deliberate reasoning. \Cref{fig.mf_method} illustrates our data curation pipeline, and \cref{fig.mf_skills} provides examples from our curated dataset.

We begin by collecting full-length songs from diverse cultures ($\sim$3M in total), as shown in ~\Cref{fig:donut}, thereby moving beyond the short, Western instrumental clips that dominate prior datasets. As shown in \cref{fig.mf_method}, our pipeline consists of four stages:  1) \textbf{Initial caption synthesis:} Generate short, surface-level captions for 30s segments using frontier music models to minimize hallucinations, 2) \textbf{Metadata extraction:} We apply conventional MIR tools, including \texttt{madmom}~\citep{böck2016madmomnewpythonaudio} (beat), \texttt{essentia}~\citep{10.1145/2502081.2502229} (key), \texttt{Chordino}~\citep{mauch2010approximate} (chords), \texttt{Parakeet}~\citep{nvidia_parakeet_tdt_0.6b_v3} (lyrics)--to provide reliable low-level attributes and, 3) \textbf{Caption \& QA creation:} Using metadata and initial captions, we prompt an LLM (with music-theory grounding) to produce detailed, multi-aspect captions covering several aspects including: a) low-level information (tempo, BPM, keys), b) instrumentation \& production, c) lyrics, and lyrical themes (including structural segmentations such as verses, choruses, and bridges), d) song structure \& dynamics e) theoretical insight (\eg, chord transitions and harmonic movements) and f) overall mood \& context. Our final captions have an average of 451.65 words. For QA, we analyze skill gaps in AF3 using benchmarks such as MMAU~\citep{sakshi2024mmau}, MMAU-Pro~\citep{kumar2025mmauprochallengingcomprehensivebenchmark}, MuChoMusic~\citep{weck2024muchomusic}, MusicCaps~\citep{agostinelli2023musiclmgeneratingmusictext}, MusicQA~\citep{li2022learning}, and NSynth~\citep{engel2017neural}, then generate novel QA targeting five skills: a) Temporal understanding, b) Attribute identification, c) Harmonic \& theoretical analysis, d) Lyric and vocal grounding, e) Comparative and structural reasoning. This approach also mitigates distribution gaps, \eg~ instrument identification in complex, multi-layered songs rather than isolated clips, and culture–skills gaps, \eg~identifying ragas in Indian music or polyrhythms in African drumming, which prior datasets neglect as they were dominated by Western music. 4) \textbf{Quality filtering:} A frontier MLLM is used to verify and retain only high-quality captions and QAs. The final dataset contains $\sim$5.2M examples ($\sim$3.4M captions and $\sim$1.8M QAs).  

Beyond curating new data, we also refine existing music datasets such as MSD~\citep{Bertin-Mahieux2011}, Music4All~\citep{geiger2025music4allaamultimodaldataset}, and the music subset of AudioSkills-XL~\citep{goel2025audioflamingo3advancing}. We (a) rewrite captions to add lyrical themes, vocal attributes, and correct mislabels of tempo, key, and timbre using metadata, and (b) reframe MCQ-style questions to reduce language priors and guessing, an issue highlighted in recent benchmarks (MMAU-Pro~\citep{kumar2025mmauprochallengingcomprehensivebenchmark}, RUListening~\citep{zang2025you}). We cluster Q\&As by trait and rephrase them with metadata to require genuine auditory perception. We provide examples below:

\begin{mdframed}[linewidth=1pt, linecolor=black, leftmargin=1pt, rightmargin=1pt, innerleftmargin=10pt, innerrightmargin=10pt, innertopmargin=4pt, innerbottommargin=2pt, backgroundcolor=gray!20, roundcorner=5pt]
\textit{\textbf{Existing caption (from AudioSkills-MSD):}} This is an upbeat 1980s pop-rock track with a danceable 4/4 beat around 140 BPM, featuring bright guitar riffs and melodic synthesizers. The song carries an energetic and catchy feel, blending indie rock elements with disco-inspired rhythms, typical of early 80s production.

\textit{\textbf{Our modified caption:}} This upbeat 1980s pop‑rock track in B minor rides a danceable 4/4 beat around 120 BPM, driven by bright electric guitars, shimmering synths, and lively drums. Its catchy, energetic melody blends indie and disco influences, $\cdots$ The lyrics add a layer of intimate storytelling, weaving lines about $\cdots$ moments of fleeting connection and dreamy recollection (“I close my eyes and count to ten $\cdots$ we were strangers a moment ago”).

\textit{\textbf{Existing QA (from AudioSkills-MSD):}} What genre does this track? Choose one among the following options: (A) Jazz (B) Classical (C) Rock (D) Spoken Word (``Spoken Word'' stands out as a unique option among all other options, which are music genres)

\textit{\textbf{Additional plausible distractors options added to modified QA:}} (E) Audiobook narration excerpt (F) Rap a cappella (no beat) (G) Podcast monologue intro (H) Documentary voice-over bed (I) Theatrical monologue with ambiance (J) Spokenword/poetryy

\end{mdframed}

\noindent \textbf{Training Methodology.} We begin with the improved base model derived from re-training Audio Flamingo 3, as outlined in \cref{sec.improved_af3}. This model is subsequently fine-tuned on the proposed MF-Skills dataset, the improved QA datasets described above, and other music datasets derived from the training mix of Audio Flamingo 3~\citep{goel2025audioflamingo3advancing}. Inspired by our study in Appendix~\ref{sec.linear_probing}, we also incorporate data for learning low-level music properties, such as chords, keys, and BPM. Dataset details in Appendix~\ref{sec.music_flamingo_training_data}

We also encounter two primary limitations in training the model. First, the Audio Flamingo 3 backbone supports a maximum context length of 8,192 tokens and $\sim$10 minutes of audio, whereas our curated datasets predominantly contain much longer captions and full-length songs up to 20 minutes. We extend the context length to $\sim$24k tokens, and adopt fully sharded training to handle the increased memory requirements. Second, music understanding requires fine-grained temporal perception, including chord progressions, tempo, key changes, and vocal dynamics. To capture these transitions, we incorporate time-aware representations into the audio encoder outputs before feeding tokens into the LLM. Specifically, we employ Rotary Time Embeddings (RoTE)~\citep{goel2024omcatomnicontextaware}, which define the rotation angle $\theta$ using absolute timestamps rather than token indices. Unlike standard RoPE, where the rotation angle $\theta$ depends on the token index $i$ as $\theta \gets -i \cdot 2\pi$, RoTE defines $\theta$ using the token’s absolute timestamp $\tau_i$: $\theta \gets -\tau_i \cdot 2\pi$. For audio tokens produced at a fixed stride of 40ms~\citep{radford2022whisper,goel2025audioflamingo3advancing}, we interpolate discrete time positions $\tau_i$ and feed them into the RoTE module to obtain lightweight, temporally grounded representations.

\subsection{Post-training with Reinforcement Learning}

While prior music understanding tasks rarely required reasoning, our formulation explicitly demands it. For example, generating a caption in \cref{fig.mf_skills} requires the model to progressively connect surface properties (tempo, key) with higher-level structures (harmony, form, production, lyrics) and then articulate them as a coherent musical narrative—a process that is non-trivial even for trained musicians. To enable this, we introduce a post-training stage beyond large-scale SFT that strengthens Music Flamingo’s reasoning abilities. First, we construct \textbf{MF-Think}, a high-quality Chain-of-Thought (CoT) dataset used for cold-start reasoning. We then fine-tune with MF-Think before applying GRPO with custom-designed rewards, encouraging explicit step-by-step reasoning.  

\noindent \textbf{MF-Think.}  We begin with a diverse, high-quality subset of MF-Skills. Since not all QAs demand deep reasoning, we sub-sample the most challenging examples by prompting \texttt{gpt-oss-120b} with both the audio and QA (prompt in Appendix~\ref{sec.prompts}). \textbf{CoT Generation.} For each selected QA or caption, we prompt \texttt{gpt-oss-120b} with metadata from MF-Skills to generate long, theory-grounded reasoning chains. Prompts include constraints on length, grounding to music theory, and exemplar demonstrations. \textbf{Quality Filtering.} Each reasoning chain is segmented into smaller steps, which are fact-checked using our post-SFT MF (Yes/No verification against the audio). We rewrite chains with minor errors and discard those with $>$30\% incorrect steps. The final dataset contains $\approx$176k CoT examples, including $\approx$117k QA and $\approx$59k captioning samples, providing a rich foundation for reasoning-enhanced training.

\noindent
\textbf{Supervised Fine-Tuning with MF-Think. } 
To equip the model with advanced reasoning capabilities, we first perform SFT of the music foundation model on our curated MF-Think dataset. During this stage, we append the prompt: \texttt{Output the thinking process in <think> </think> and final answer in <answer> </answer>} to the original questions in the \textbf{MF-Think} dataset, to encourage the model to explicitly generate reasoning chains within the \texttt{<think>} \texttt{</think>} tags and the final answer within the \texttt{<answer>} \texttt{</answer>} tags.
This process instills structured reasoning for both the question-answering and caption-generation tasks. This SFT stage acts as an initial warm-up phase, effectively priming the model for subsequent reinforcement learning (RL) fine-tuning.

\noindent
\textbf{GRPO for Music reasoning and understanding. } Building on the advancements in the GRPO algorithm, we adhere to the standard GRPO algorithm to train our model. GRPO obviates the need for an additional value function and uses the average reward of multiple sampled outputs for the same question to estimate the advantage. For each given question $q$, the policy model generates a group of candidate responses $\{o_1, o_2, \ldots, o_G\}$ from the old policy $\pi_{\theta_{\text{old}}}$
along with their corresponding rewards $\{r_1, r_2, \ldots, r_G\}$ which are computed using rule-based reward functions (\eg, format and accuracy). The model $\pi_{\theta}$ is subsequently optimized using the following objective function: 

\begin{align}
\mathcal{J}(\theta)\!\! = \!\!\mathbb{E}_{q,\{o_i\}} \!\!\left[\! \tfrac{1}{G} \!\sum_{i=1}^{G} \!\Big(\!\! \min \!\Big(\! \tfrac{\pi_\theta(o_i|q)}{\pi_{\theta_{\text{old}}}(o_i|q)} A_i, \, \text{clip}\!\big(\tfrac{\pi_\theta(o_i|q)}{\pi_{\theta_{\text{old}}}(o_i|q)},\! 1-\epsilon, \!1+\epsilon \big) A_i \Big)\! -\! \beta D_{\text{KL}}(\pi_\theta \,\|\, \pi_{\text{ref}}) \!\Big)\! \right]\!
\label{eq:grpo_objective}
\end{align} 

where $\epsilon$ is the clipping range of the importance sampling ratio, $\beta$ is the regularization strength of the KL-penalty term that encourages the learned policy to stay close to the reference policy, and $G$ is the group size, \ie, the number of candidate responses (samples) the policy generates for each input question, which is set to $5$ in our experiments. To stabilize training, the sampled rewards are normalized to compute the advantages $A_i$ as:
$\frac{r_i - \text{mean}(\{r_1, r_2, \ldots, r_G\})}{\text{std}(\{r_1, r_2, \ldots, r_G\})}$
Next, we discuss the custom reward functions we designed for GRPO training, which play a pivotal role in optimization. 

\noindent
\textbf{Format Reward. }
In order to encourage the model to generate outputs that adhere to the prescribed response format, we use the standard regex-based format reward~\citep{deepseekai2025deepseekr1incentivizingreasoningcapability}. Specifically, the model is instructed to produce reasoning traces enclosed within  \texttt{<think>} \texttt{</think>} tags, followed by the final answer enclosed within \texttt{<answer>} \texttt{</answer>} tags.   If the output strictly follows the required tag structure, the model gets a reward of 1 else 0. This binary reward function ensures that the model learns to consistently produce well-structured responses.

\noindent
\textbf{Accuracy Reward. }
For question-answering (QA) tasks, we employ the \textit{accuracy reward} to encourage the model to generate accurate final answers. Given a question with the corresponding ground-truth answer, the model generates a candidate output $o_i$, where the final answer is extracted from within the \texttt{<answer>} \texttt{</answer>} tags.  
The accuracy reward directly matches the normalized predicted and ground truth answers, enforcing strict answer correctness. 

\textbf{Structured Thinking Reward. } 
For caption generation tasks, the standard accuracy reward cannot be directly applied due to the long and open-ended nature of the generated captions. 
To address this, we design a custom reward function that evaluates generated captions against structured ground-truth metadata. 
To achieve this, we first generate ground-truth structured metadata as shown below using \texttt{gpt-oss-120b}~\citep{openai2025gptoss120bgptoss20bmodel} for the subset of the captions in the MF-Skills dataset as follows:

\begin{mdframed}[linewidth=1pt, linecolor=black, leftmargin=1pt, rightmargin=1pt, innerleftmargin=10pt, innerrightmargin=10pt, innertopmargin=4pt, innerbottommargin=2pt, backgroundcolor=gray!20, roundcorner=5pt]
\{``\textbf{Genre}'': Americana, ``\textbf{BPM}'': 125, ``\textbf{Key}'': G minor, ``\textbf{Meter}'': 4/4, ``\textbf{Structure}'': Intro, Verse, Verse, Bridge, Solo, Chorus, Outro, ``\textbf{Instruments}'': fingerstyle acoustic guitar, banjo, mandolin, spoken-word voice, ``\textbf{Vocal Character}'': male spoken-word, deep resonant timbre, clear/deliberate, light reverb, ``\textbf{Lyric Themes}'': forgiveness, humility, spiritual prayer, desert frontier, betrayal, outlaw narrative, ``\textbf{Theory}'': G minor center; modal interchange with relative major; F\#aug → Eb6 → D7 resolution; Gmaj7/G7 brighten prayer sections, ``\textbf{Mix Notes}'': high-fidelity organic; wide natural stereo panning; minimal reverb; warm, clear, light compression; close-mic intimacy, ``\textbf{Dynamics}'': bridge increases harmonic rhythm/urgency.\}
\end{mdframed}

The structured thinking reward function computes a string match for each answer in the category of the structured ground-truth metadata (\eg~Genre, Subgenre, BPM etc.) and the generated caption. The total reward is obtained by normalizing the number of matching words by the total number of metadata categories.

The overall reward function used in GRPO training integrates the format reward with the accuracy reward for the data with question-answer subset, while for the caption subset of data, it combines the format reward with the structured reasoning reward.

% \vspace{-2mm}

\section{Experiments}
\label{sec:experiments}
\vspace{-1mm}
\begin{table}[!t]
\centering
\caption{\small Comparison of Music Flamingo (w/ GRPO) with other LALMs on various benchmarks (WER ↓ (Word Error Rate), ACC ↑ (Accuracy), Score (1-10) ↑ and GPT5 ↑ (GPT evaluation)). We report scores for only the top-performing prior LALM. We highlight \textcolor{closedGray}{closed source}, \textcolor{qwenPurple}{open weights}, and \textcolor{nvidiaGreen}{open source} models.}
\vspace{-2mm}
\resizebox{\linewidth}{!}{
\begin{tabular}{llccc}
\toprule
\textbf{Task} & \textbf{Dataset} & \textbf{Model} & \textbf{Metrics} & \textbf{Results} \\
\midrule

\multirow{14}{*}{\shortstack[c]{\textbf{Music QA} \\ \textbf{and} \\\textbf{Reasoning}}} 
& \multirow{2}{*}{\shortstack[l]{\textbf{MMAU (Music)} \\ \textit{full-test \textbar{} test-mini} }} & \textcolor{nvidiaGreen}{Audio Flamingo 3} & \multirow{3}{*}{ACC ↑} & 73.95 \textbar{}  74.47\\
& & \textcolor{nvidiaGreen}{Music Flamingo} & & \textbf{76.83} \textbar{} \textbf{76.35} \\ \cmidrule{2-5}

& \multirow{2}{*}{\shortstack[l]{\textbf{MMAU-Pro-Music} }}
& \textcolor{closedGray}{Gemini-2.5 Flash} & \multirow{3}{*}{ACC ↑} & 64.90 \\
& & \textcolor{nvidiaGreen}{Music Flamingo} & &  \textbf{65.60} \\ \cmidrule{2-5}

& \multirow{2}{*}{\shortstack[l]{\textbf{MuChoMusic } }}
& \textcolor{qwenPurple}{Qwen3-O} & \multirow{3}{*}{ACC ↑} & 52.10 \\
& & \textcolor{nvidiaGreen}{Music Flamingo} & & \textbf{74.58}  \\ \cmidrule{2-5}

& \multirow{2}{*}{\shortstack[l]{\textbf{MMAR (Music)} }}
& \textcolor{qwenPurple}{Qwen2.5-O} & \multirow{3}{*}{ACC ↑} & 46.12\\
& & \textcolor{nvidiaGreen}{Music Flamingo} & & \textbf{48.66}  \\\cmidrule{2-5}

& \multirow{2}{*}{\textbf{Music Instruct}} & \textcolor{nvidiaGreen}{Audio Flamingo 3} & \multirow{2}{*}{GPT5 ↑} &  92.7\\
& & \textcolor{nvidiaGreen}{Music Flamingo} & & \textbf{97.1}  \\ \cmidrule{2-5}

& \multirow{2}{*}{\textbf{Music AVQA}} & \textcolor{nvidiaGreen}{Audio Flamingo 3} & \multirow{2}{*}{ACC ↑} &  \textbf{76.7}\\
& & \textcolor{nvidiaGreen}{Music Flamingo} & &  73.6 \\ \cmidrule{2-5}

& \multirow{2}{*}{\shortstack[l]{\textbf{SongCaps (Ours)} \\ \textit{Human \textbar{} GPT5-Coverage \textbar{} GPT5-Correctness} }} & \textcolor{nvidiaGreen}{Audio Flamingo 3} & \multirow{2}{*}{Score ↑} &  6.5 \textbar{} 6.7 \textbar{} 6.2 \\
& & \textcolor{nvidiaGreen}{Music Flamingo} & &  \textbf{8.3} \textbar{} \textbf{8.8} \textbar{} \textbf{8.0} \\ 
\midrule

\multirow{8}{*}{\shortstack[c]{\textbf{Music} \\ \textbf{Information} \\\textbf{Retrieval}}} 

& \multirow{2}{*}{\shortstack[l]{\textbf{NSynth} \\ \textit{Source \textbar{} Instrument} }} & \textcolor{nvidiaGreen}{Audio Flamingo 3} & \multirow{2}{*}{ACC ↑} & 65.5 \textbar{} 78.9 \\
& & \textcolor{nvidiaGreen}{Music Flamingo} & &  \textbf{75.89} \textbar{} \textbf{80.76}\\ \cmidrule{2-5}

& \multirow{2}{*}{\shortstack[l]{\textbf{GTZAN} \\ \textit{Genre} }} & \textcolor{nvidiaGreen}{Pengi} & \multirow{2}{*}{ACC ↑} &  80.00\\
& & \textcolor{nvidiaGreen}{Music Flamingo} & & \textbf{84.45} \\ \cmidrule{2-5}

& \multirow{2}{*}{\shortstack[l]{\textbf{Medley-Solos-DB} \\ \textit{Instrument} }} & \textcolor{nvidiaGreen}{Audio Flamingo 2} & \multirow{2}{*}{ACC ↑} &  85.80 \\
& & \textcolor{nvidiaGreen}{Music Flamingo} & &  \textbf{90.86} \\ \cmidrule{2-5}

& \multirow{2}{*}{\textbf{MusicCaps}} & \textcolor{qwenPurple}{Qwen3-O} & \multirow{2}{*}{GPT5 ↑} & 7.2 \\
& & \textcolor{nvidiaGreen}{Music Flamingo} & & \textbf{8.8} \\ 

\midrule

\multirow{4}{*}{\shortstack[c]{\textbf{Lyrics} \\ \textbf{Transcription}}} 

& \multirow{2}{*}{\shortstack[l]{\textbf{Opencpop} \\ \textit{Chinese} }}& \textcolor{closedGray}{GPT-4o} \textbar{} \textcolor{qwenPurple}{Qwen2.5-O} & \multirow{2}{*}{WER ↓} &  53.7 \textbar{} 55.7\\
& & \textcolor{nvidiaGreen}{Music Flamingo} & &  \textbf{12.9} \\ \cmidrule{2-5}

& \multirow{2}{*}{\shortstack[l]{\textbf{MUSDB18} \\ \textit{English} }} & \textcolor{closedGray}{GPT-4o} \textbar{} \textcolor{qwenPurple}{Qwen2.5-O} & \multirow{2}{*}{WER ↓} &  32.7 \textbar{} 68.7 \\
& & \textcolor{nvidiaGreen}{Music Flamingo} & &  \textbf{19.6} \\

\bottomrule
\end{tabular}}
\vspace{-4mm}
\label{tab:main_results}
\end{table}

\noindent
\textbf{Experimental Setup. }We train Music Flamingo on 128 NVIDIA A100 GPUs (80GB). Details on batch size, learning rates, and optimizers for each stage of training are in Appendix~\ref{sec.music_flamingo_training_details}. 

\noindent \textbf{Baselines.} We evaluate our model against recent SOTA LALMs, including GAMA~\citep{ghosh2024gama}, Audio Flamingo~\citep{kong2024audioflamingo}, Audio Flamingo 2~\citep{kong2025audioflamingo2}, Audio Flamingo 3~\citep{goel2025audioflamingo3advancing}, Qwen-Audio~\citep{chu2023qwenaudio}, Qwen2-Audio~\citep{chu2024qwenaudio2}, Qwen2-Audio-Instruct, Qwen2.5-Omni~\citep{xu2025qwen2}, R1-AQA~\citep{li2025reinforcement}, Pengi~\citep{deshmukh2023pengi}, Phi-4-mm~\citep{abouelenin2025phi}, Baichun Audio~\citep{li2025baichuan}, Step-Audio-Chat~\citep{huang2025step}, LTU~\citep{gong2023ltu}, LTU-AS~\citep{gong2023ltu-as}, SALMONN~\citep{tang2024salmonngenerichearingabilities}, AudioGPT~\citep{huang2023audiogpt}, and Gemini (2.0 Flash, 1.5 Pro, 2.5 Flash and 2.5 Pro)~\citep{team2023gemini}, as well as GPT-4o-audio~\citep{hurst2024gpt}. For Table~\ref{tab:main_results}, we only compare against open LALMs. All results reported in the tables correspond to the best-performing model. 

\noindent \textbf{Evaluation Datasets.} We evaluate AF3 across a broad set of benchmarks spanning music information retrieval (MIR), question answering, lyrics transcription, reasoning, and our proposed dataset \textbf{SongCaps}. SongCaps consists of 1,000 culturally diverse songs curated to assess captioning capabilities across multiple dimensions (see Section~\ref{sec:bulding_music}). Rather than relying on lexical overlap metrics, we evaluate captions using human-expert judgments and LLM-as-a-judge assessments. For MIR, we use NSynth (Source and Instrument)~\citep{engel2017neural}, MusicCaps~\citep{agostinelli2023musiclm}, Medley-Solos-DB (instrument classification)~\citep{lostanlen_2019_3464194}, and GTZAN (genre classification)~\citep{tzanetakis2002gtzan}. For QA and reasoning, we include MusicAVQA~\citep{li2022learning}, Music Instruct~\citep{deng2024musilingobridgingmusictext}, MMAU (v05.15.25)~\citep{sakshi2024mmau}, MMAU-Pro~\citep{kumar2025mmauprochallengingcomprehensivebenchmark}, MuChoMusic (perceptual version)~\citep{zang2025you,weck2024muchomusic}, and MMAR~\citep{ma2025mmarchallengingbenchmarkdeep}. For lyrics transcription, we evaluate on Opencpop~\citep{wang2022opencpophighqualityopensource} -- a dataset for chinese songs and MUSDB18 Lyrics~\citep{musdb18-hq} -- a dataset for English songs. \textit{We acknowledge the existence of numerous other MIR baselines and benchmarks, as MIR encompasses a broad range of tasks. For the scope of this paper, however, we restrict our comparisons to LALMs and the benchmarks most commonly used in the LALM literature. We encourage the community to further expand evaluations to a wider set of MIR baselines in future work.}

\noindent \textbf{Music Understanding and Reasoning Evaluation.} Table~\ref{tab:main_results} shows that Music Flamingo consistently sets the bar across music QA, reasoning, MIR, and lyrics transcription benchmarks. On MMAU-Music, it reaches a competitive 76.83 accuracy, surpassing both closed and open-source models. The gap widens on the tougher MMAU-Pro-Music and MuChoMusic benchmarks, where Music Flamingo scores 65.6 and 74.58, respectively, a clear evidence of its robustness on complex datasets. Without reinforcement learning fine-tuning with thinking traces, performance drops to 63.9 and 69.5 respectively, highlighting the value of step-by-step reasoning and exploration.
In MIR tasks, Music Flamingo continues to dominate: on NSynth, it achieves 80.76\% accuracy in instrument recognition, and on Medley Solos DB, it reaches 90.86\% for fine-grained instrument classification. It also delivers a significantly lower WER on Chinese and English lyrics transcription than both open and closed-source LALMs.
These results establish Music Flamingo as the most capable model to date for detailed music understanding and reasoning.

On our proposed SongCaps benchmark, designed to evaluate music captioning, human raters scored the model outputs on a scale of 1-10. Music Flamingo achieves a high rating of 8.3 outperforming Audio Flamingo 3. Furthermore, we evaluate the captions using LLM-as-a-judge measuring both \textit{correctness} and \textit{coverage} of the caption. Music Flamingo achieves 8.0 for correctness and 8.8 for coverage, outperforming Audio Flamingo 3. These results highlight that Music Flamingo not only excels on structured QA and recognition tasks but also produces richer, more faithful natural language descriptions of music.

\noindent \textbf{Qualitative Evaluation. }
We perform a thorough qualitative evaluation of Music Flamingo’s outputs, assessed by trained music experts, in comparison with state-of-the-art LALMs in this domain. Due to space constraints, we refer the readers to Appendix~\ref{sec.user_study} for analysis on songs of varying genres and popularity, and additionally analysis of songs from different cultures in Appendix~\ref{sec.user_study_cultures}.

 \section{Conclusion, Limitations and Future Work}
\label{sec:discussion}
\vspace{-1mm}

We introduced Music Flamingo, a large audio–language model designed to advance music understanding. By curating MF-Skills and MF-Think, we scale beyond short, instrumental clips to full-length, multi-cultural songs with layered annotations, and incorporate chain-of-thought reasoning for richer music analysis. Through improved pretraining, fine-tuning, and post-training with reinforcement learning, Music Flamingo achieves SOTA results across diverse music understanding and reasoning benchmarks. Beyond empirical gains, it demonstrates how models can move from surface-level recognition toward layered, human-like perception of songs.

Music Flamingo still has a few limitations, including: (i) limited understanding of underrepresented or skewed cultural traditions, highlighting the need to expand training data across more diverse global music; (ii) gaps in specialized tasks, such as fine-grained piano technique recognition and other instrument-specific skills; and (iii) the need to broaden coverage across additional musical skills to achieve more comprehensive understanding.

\bibliography{iclr2026_conference}
\bibliographystyle{iclr2026_conference}

\appendix
\section*{Appendix}
\label{sec:appendix}
\section{Ethics Statement}
This work studies audio, music, and singing‐voice understanding across culturally diverse material. Our experiments rely on publicly available datasets and/or content licensed for research use. We do not release copyrighted audio, stems, or lyrics; any examples used for qualitative illustration are either (i) already distributed by the originating dataset under a research‐permissive license, or (ii) replaced by non-copyrightable descriptors (e.g., metadata, short transcriptions for analysis) when licenses are restrictive. No personally identifying information is collected, and no human-subjects experiments were conducted; institutional review board (IRB) approval was therefore not required.

\textbf{Cultural representation and bias.} Music corpora are uneven across regions, languages, and genres. Such imbalance can yield biased estimates or degrade performance on underrepresented traditions. We mitigate this by (a) documenting dataset composition and selection criteria, (b) emphasizing vocal and multicultural material in evaluation, and (c) reporting known limitations. We encourage downstream users to avoid normative claims about “quality” across cultures and to treat our benchmarks as descriptive rather than prescriptive.

\textbf{Copyright and content ownership.} Models trained on musical recordings risk reproducing protected content. We do not deploy or evaluate generative audio synthesis; our outputs are textual (QA, captions, reasoning traces). We avoid releasing any asset that could enable reconstruction of substantial portions of copyrighted works and provide guidance for filtering long verbatim lyric reproduction in evaluation outputs.

\textbf{Privacy and safety.} Singing voices may implicitly encode sensitive traits. We use only public research datasets and focus on musical attributes (rhythm, harmony, timbre, structure) rather than identifying individuals. Potential misuse includes intrusive listener profiling or surveillance via audio analysis; to discourage these, we release only research artifacts (documentation, evaluation protocols, non-identifying metadata) and clearly scope permitted use in licenses when possible.

\section{Reproducibility Statement}
We provide all details needed to reproduce our results within the paper and appendix: dataset sources and splits; audio preprocessing (sampling rates, normalization, chunking/segment lengths); model architectures and parameter counts; training schedules (optimizers, learning-rate policies, batch sizes, gradient clipping), GRPO/post-training settings and reward definitions; inference settings (temperature, decoding constraints); and exact evaluation protocols and metrics for every benchmark. We report hardware used where applicable, and mean$\pm$std for repeated trials. We will release code, checkpoints and data upon acceptance.

\section{Music Flamingo Training Datasets}
\label{sec.music_flamingo_training_data}

Table~\ref{tab:sft_datasets_app} summarizes all datasets used to train Music Flamingo, including total hours, number of audio-QA pairs, and the number of epochs (passes over the dataset) used at each training stage. 
Similar to ~\citep{kong2025audioflamingo2, goel2025audioflamingo3advancing}, we convert all foundational datasets (captioning, classification, etc.) into QA formats, using the same set of prompts for each task mentioned in ~\citep{kong2025audioflamingo2, goel2025audioflamingo3advancing}.

\begin{table*}[h]
    \centering
    \caption{\small List of fine pre-training and fine-tuning datasets together with their training composition.}
    \resizebox{\textwidth}{!}{
    \begin{tabular}{lcccccc}
    \toprule
Dataset & Hours  & Num. Pairs & AF3-St. 3 & MF-SFT & MF-Warmup & MF-GRPO \\ \midrule
        AF3-training mix~\citep{goel2025audioflamingo3advancing} & - & 30M & 1.0 & - & - & - \\
        MF-Skills (Ours) & - & 3M & - & 2.0 & - & 1.0 \\
        MF-Think (Ours)& - & 176k & - & - & $1.0$& $1.0$\\
        MusicBench~\citep{melechovsky2023mustango} & 115.5 hrs & 686k & 1.0 & 1.0& - & -\\
        Mu-LLAMA~\citep{liu2024music} & 62.9 hrs & 70k & 1.0 & 1.0& - & -\\
        MusicAVQA\textsubscript{audio-only}~\citep{li2022learning} & 77.1 hrs & 5.7K & 1.0 & 1.0& - & -\\
        MusicQA~\citep{ouyang2025mqad} & 62.9 hrs & 70K & 1.0 & 1.0 & -& - \\
        LP-MusicCaps\textsubscript{MSD}~\citep{doh2023lp} & 5805.7 hrs & 1331.8K & 1.0 & 1.0& - & -\\
        LP-MusicCaps\textsubscript{MTT}~\citep{doh2023lp} & 126.4 hrs & 46.9K & 1.0 & 1.0& - & -\\
        LP-MusicCaps\textsubscript{MC}~\citep{doh2023lp} & 7.4 hrs & 7.9K & 1.0 & 1.0& - & -\\
        MusicCaps~\citep{agostinelli2023musiclmgeneratingmusictext} & 7.4 hrs & 2.6K & 1.0 & 1.0& - & -\\
        NSynth~\citep{engel2017neural} & 321.3 hrs & 289.2K & 1.0 & 1.0& - & -\\
        MusDB-HQ~\citep{rafii2017musdb18} & 29.1 hrs & 10.2K & 1.0 & 1.0& - & -\\
        FMA~\citep{defferrard2016fma} & 860.7 hrs & 104.2K & 1.0 & 1.0& - & -\\
    
        % MusicAVQA\textsubscript{audio-visual}~\citep{li2022learning} & 142.4 hrs & 17.9K & $1.0$ & $6.0$& $6.0$& -& -\\
        % MTG-Jamendo & 3768.9 hrs & 55.6K & $1.0$ & -\\
        Music4All Captions (ours)& 910.5 hrs & 109k  & 1.0 & 1.0& - & -\\
        Music4All QA (ours)& 1505.7 hrs & 180k & 1.0 & 1.0& - & - \\
        MSD Captions (ours)& 15449.9 hrs & 1.4M & 1.0 & 1.0& - & -\\
        MSD QA (ours)& 20906.2 hrs & 935k & 1.0 & 1.0& - & - \\

        % Speech-in-Sound QA& 4833.2 hrs & 55.6K & $1.0$ & -\\
        CHIME~\citep{foster2015chime} & 342 hrs & 30k & $1.0$ & -&- & -\\
        ALI Meeting~\citep{yu2022m2meticassp2022multichannel} & 118.75 hrs & 387k & $1.0$ & -&- & -\\
        EMILIA~\citep{he2024emiliaextensivemultilingualdiverse} & 5000 hours & 1.7M & $1.0$ & -&- & -\\
        MUST~\citep{qin2025mustdatasetunifiedframework} & 500 hrs & 245k & $1.0$ & -&- & - \\
        CoVoST~\citep{wang2020covost2massivelymultilingual} & 2880 hrs & 5M & $1.0$ & -&- & -\\
        Multi-talker Switchboard~\citep{godfrey1992switchboard} & 109.9 hrs & 76.6K & $1.0$ & -&- & -\\
       
    \bottomrule
    \end{tabular}}
    
    \label{tab:sft_datasets_app}
\end{table*}

% \newpage

\section{Music Flamingo Training Details}
\label{sec.music_flamingo_training_details}
In this section, we present the training settings of our model across all stages, each with specific configurations. Details are in \cref{tab:hyperparams}.

\begin{table}[!ht]
\centering
\begin{tabular}{lcccc}
\toprule
\textbf{Settings} & \textbf{AF3-SFT} & \textbf{MF-SFT} & \textbf{MF-WarmUp} & \textbf{MF-GRPO} \\
\midrule
global batch size & 128 & 128 & 128 & 64 \\
learning rate & 1.5e-5 & 1.5e-5 & 1e-5 & 1e-6\\
learning schedule & \multicolumn{4}{c}{Cosine decay} \\
warm up ratio & \multicolumn{4}{c}{0.03}\\
weight decay & \multicolumn{4}{c}{0.0}\\
epoch & 1 & 1 & 1 & 1\\
bf16 & \checkmark & \checkmark & \checkmark & \checkmark \\
grad accumulate & \multicolumn{4}{c}{8} \\
FSDP  -- full shard & \checkmark & \checkmark & \checkmark & \checkmark \\
GPUs & \multicolumn{4}{c}{128$\times$A100} \\
\bottomrule
\end{tabular}
\caption{\small Training settings across stages.}
\label{tab:hyperparams}

\end{table}

\section{User Study on Music Flamingo}
\label{sec.user_study}
We undergo a user study with trained music experts comparing Music Flamingo to an open-source LALM--\emph{Qwen3 Omni}, and two closed-source LALMs--(\emph{GPT-4o} and \emph{Gemini 2.5 Pro}) qualitatively. To achieve this, we selected a subset of 8 songs: 4 songs in English and 4 songs in Brazilian Portuguese. From these songs, half of them are by extremely popular artists in Western music, and other half from less known artists. The following songs were used: 1) \textit{ABBA - Money Money Money}, 2) \textit{Michael Sembello - Maniac (From Flashdance)}, 3) \textit{Chandler Leighton - NO I DON T}, 4) \textit{Lø Spirit - Wild Things}, 5) \textit{Antônio Carlos Jobim - Águas De Março}, 6) \textit{Michel Teló - Ai Se Eu Te Pego}, 7) \textit{Paulinho da Viola - Apoteose Ao Samba} and 8) \textit{Ave Sangria - Seu Waldir}.

\Cref{tab:model_comparison} shows a summary of the detailed analysis comparing different musical aspects and features across models. Among the four models, Music Flamingo performs the best overall while some limitations in accurately identifying deeper context remain.

\begin{table}[!t]
\centering
\resizebox{\textwidth}{!}{%
\begin{tabular}{p{3.5cm} p{3.2cm} p{3.2cm} p{3.2cm} p{3.2cm}}
\toprule
\textbf{Aspect} & \textbf{Music Flamingo} & \textbf{Qwen3-Omni} & \textbf{GPT4o-Audio} & \textbf{Gemini-2.5 Pro} \\
\midrule
\textbf{General technical characteristics (tempo, key, time signature)} 
& Consistently outputs tempo (bpm) and key; sometimes time signatures (3 songs correctly 4/4). Mistakes usually in relative major/minor. 
& Rarely detailed; often omits tempo/key/time signature. 
& Sometimes outputs tempo (bpm usually in ballpark). Less consistent with key. 
& Sometimes outputs tempo (closest bpm for \textit{NO I DON’T}). Some key mistakes. Time signatures generally omitted. \\
\midrule
\textbf{Genre classification }
& Reasonably good, but misclassified \textit{Maniac} (electro-funk vs. dance-pop). Struggled with Brazilian music genres (occasional mismatches). 
& Superficial; often wrong with Brazilian genres. 
& Some correct, but not consistent. 
& Best at identifying genres overall, but hallucinated (\textit{ABBA} as ska cover; \textit{Wild Things} as dream pop with drums). \\
\midrule
\textbf{Emotional content \& lyrics }
& Good understanding; captures emotional context. Lacks cultural/historical nuance. 
& Similar to Music Flamingo; misses deeper cultural context. 
& More superficial understanding than others. 
& Good understanding; one small lyric mistake. Also lacks cultural/historical nuance. \\
\midrule
\textbf{Complex technical characteristics (chord progressions, structure, production)} 
& Attempts deeper detail, but sometimes inaccurate with chords/voicings/structure. Sometimes hallucinates genre-related elements. 
& Least detailed; only superficial composition/arrangement notes. 
& Gives more detail when recognizing famous songs; otherwise shallow. 
& More detailed for famous songs; hallucinations (e.g., nonexistent drums, overstated synths). Genre misclassifications cascade into wrong technical details. \\
\midrule
\textbf{General observations }
& Strong in consistent technical reporting, but accuracy varies on deeper features. 
& Shallowest outputs. 
& Relies on recognition of famous songs; may pull from text knowledge instead of audio. 
& Similar to GPT4o-Audio; detailed when recognizing famous songs. Most prone to hallucinations tied to genre assumptions. \\
\bottomrule
\end{tabular}%
}
\caption{Comparison of Music Flamingo, Qwen3-Omni, GPT4o-Audio, and Gemini 2.5 Pro across different evaluation aspects.}
\label{tab:model_comparison}
\end{table}

\newpage

% Lasha's version on the 5 songs
\section{Comparative Analysis Across Songs from Different Cultures}
\label{sec.user_study_cultures}

\begin{figure}[!ht]
    \centering
    \includegraphics[width=\linewidth]{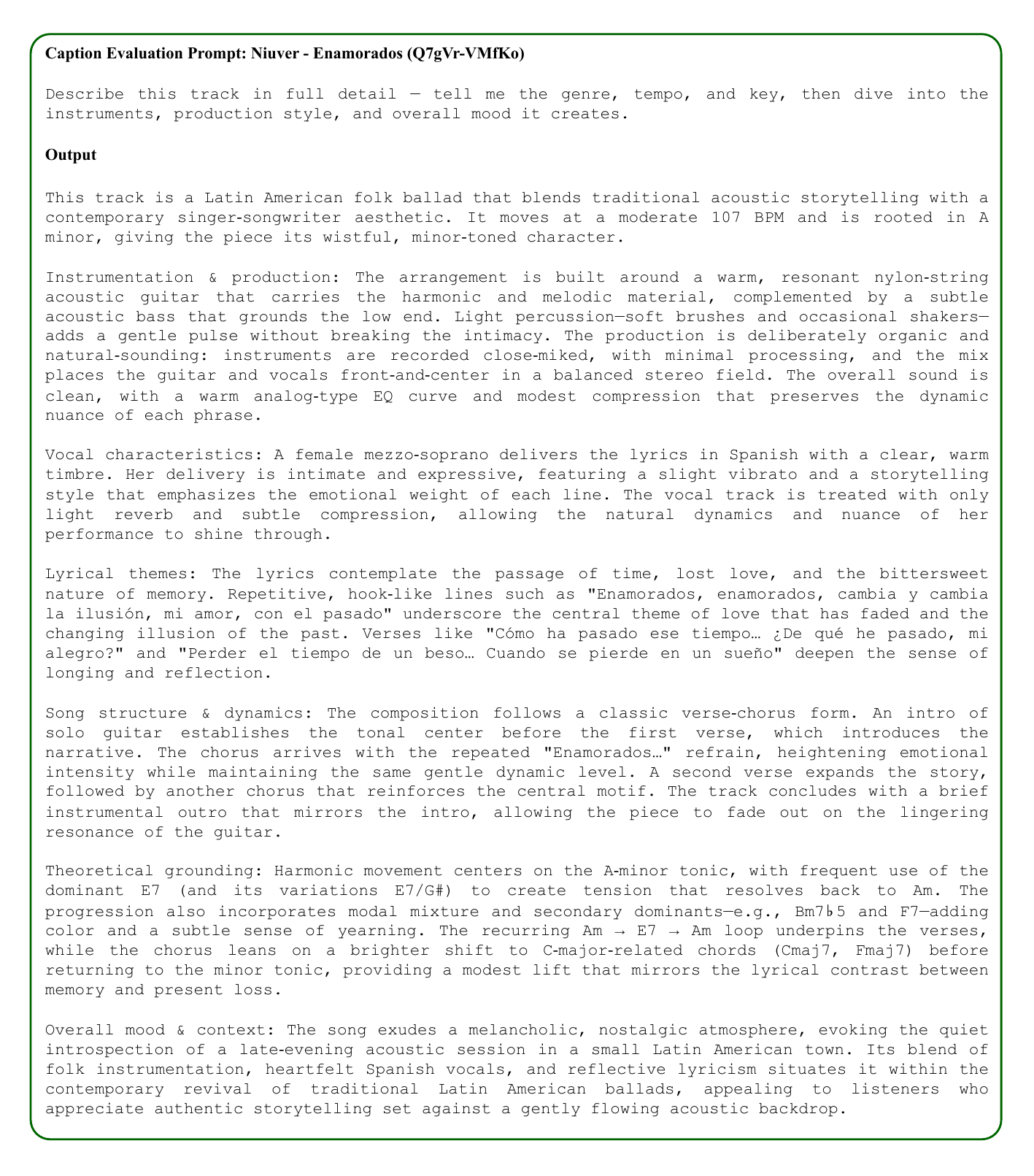}
    \caption{Caption generated by Music Flamingo on a modern Spanish song.}
    \label{fig.flamingo_latin}
    \vspace{-2mm}
\end{figure}

\begin{figure}[!ht]
    \centering
    \includegraphics[width=\linewidth]{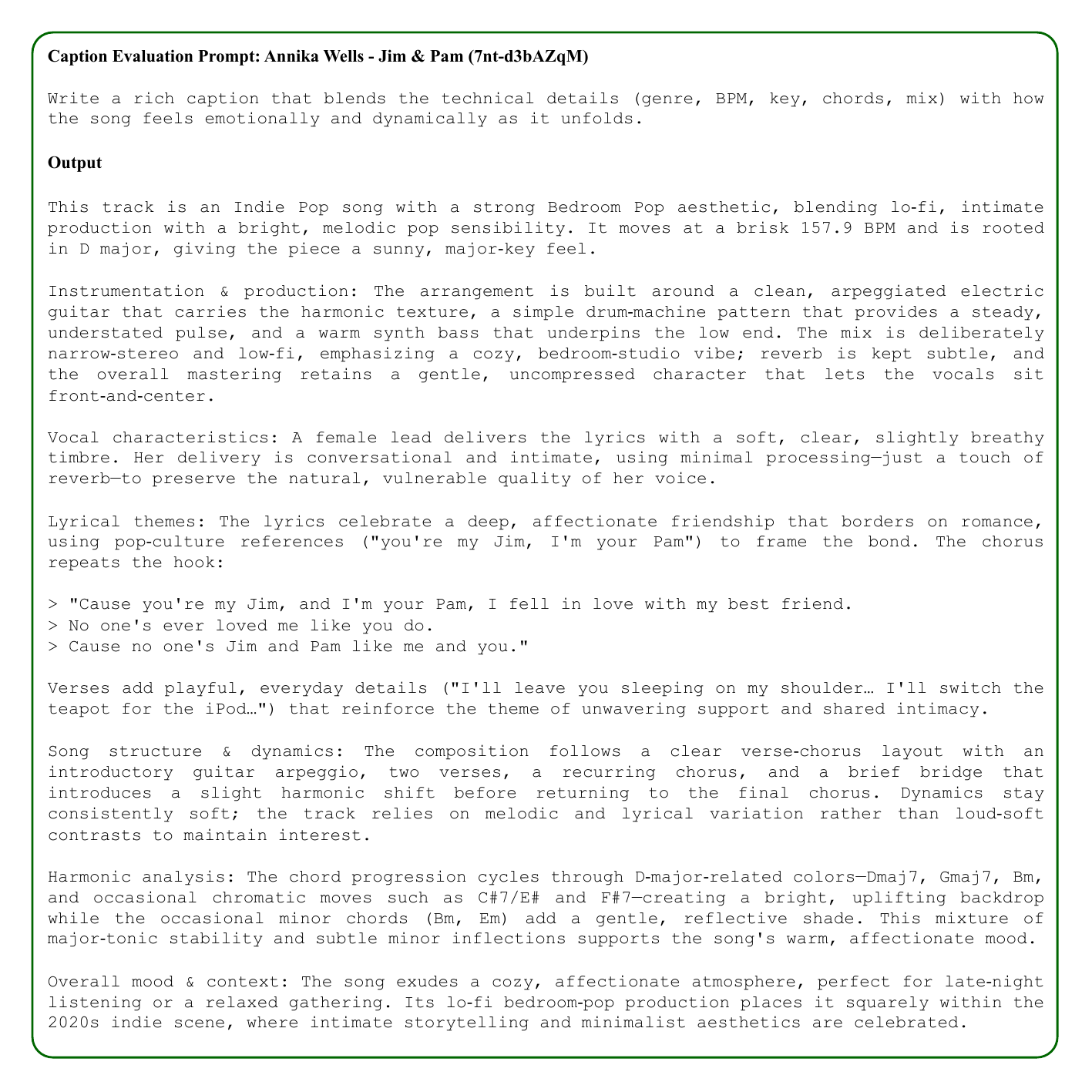}
    \caption{Caption generated by Music Flamingo on a less known American song.}
    \label{fig.flamingo_american}
    \vspace{-2mm}
\end{figure}

\begin{figure}[!ht]
    \centering
    \includegraphics[width=\linewidth]{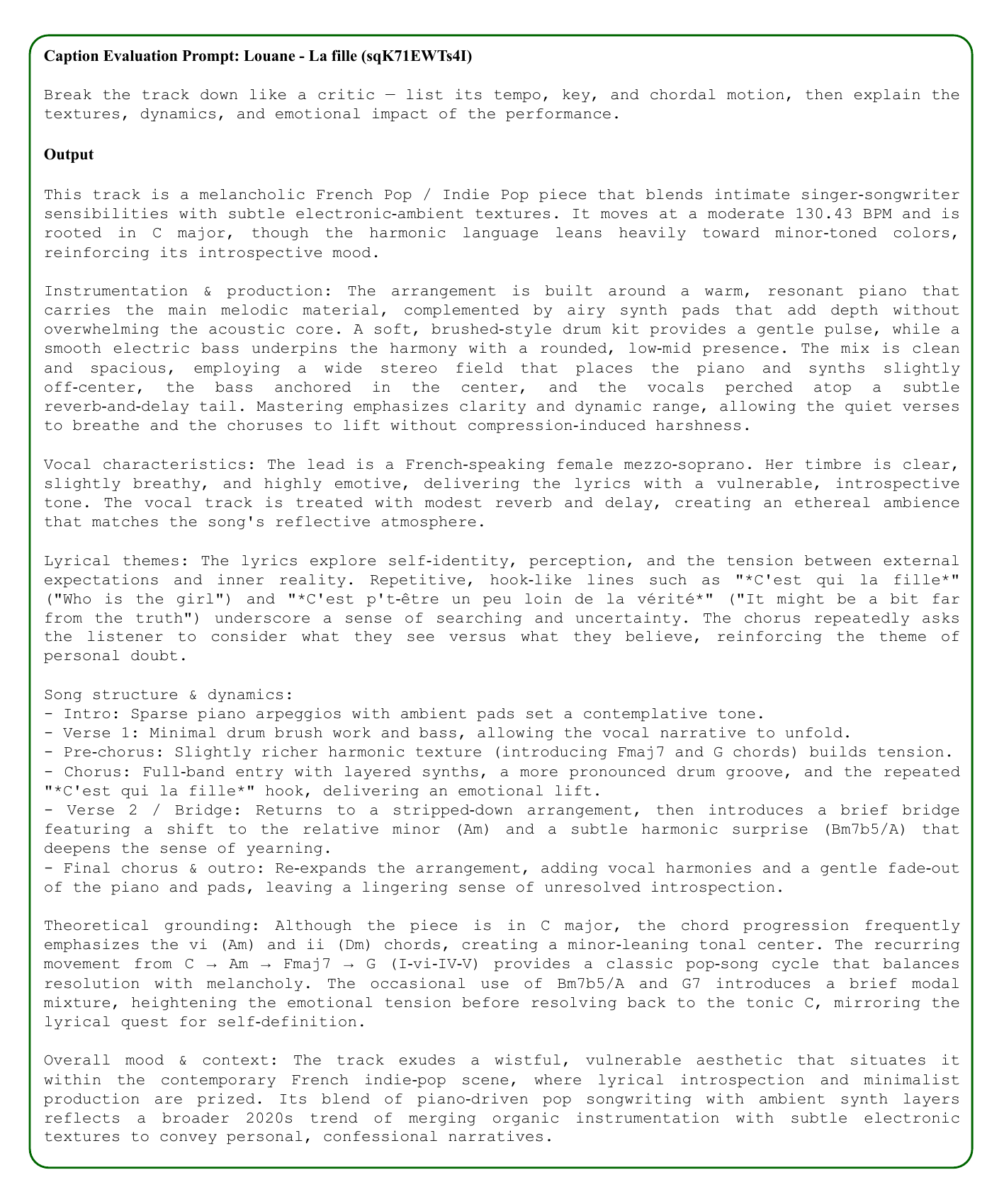}
    \caption{Caption generated by Music Flamingo on a modern French song.}
    \label{fig.flamingo_french}
    \vspace{-2mm}
\end{figure}

\begin{figure}[!ht]
    \centering
    \includegraphics[width=\linewidth]{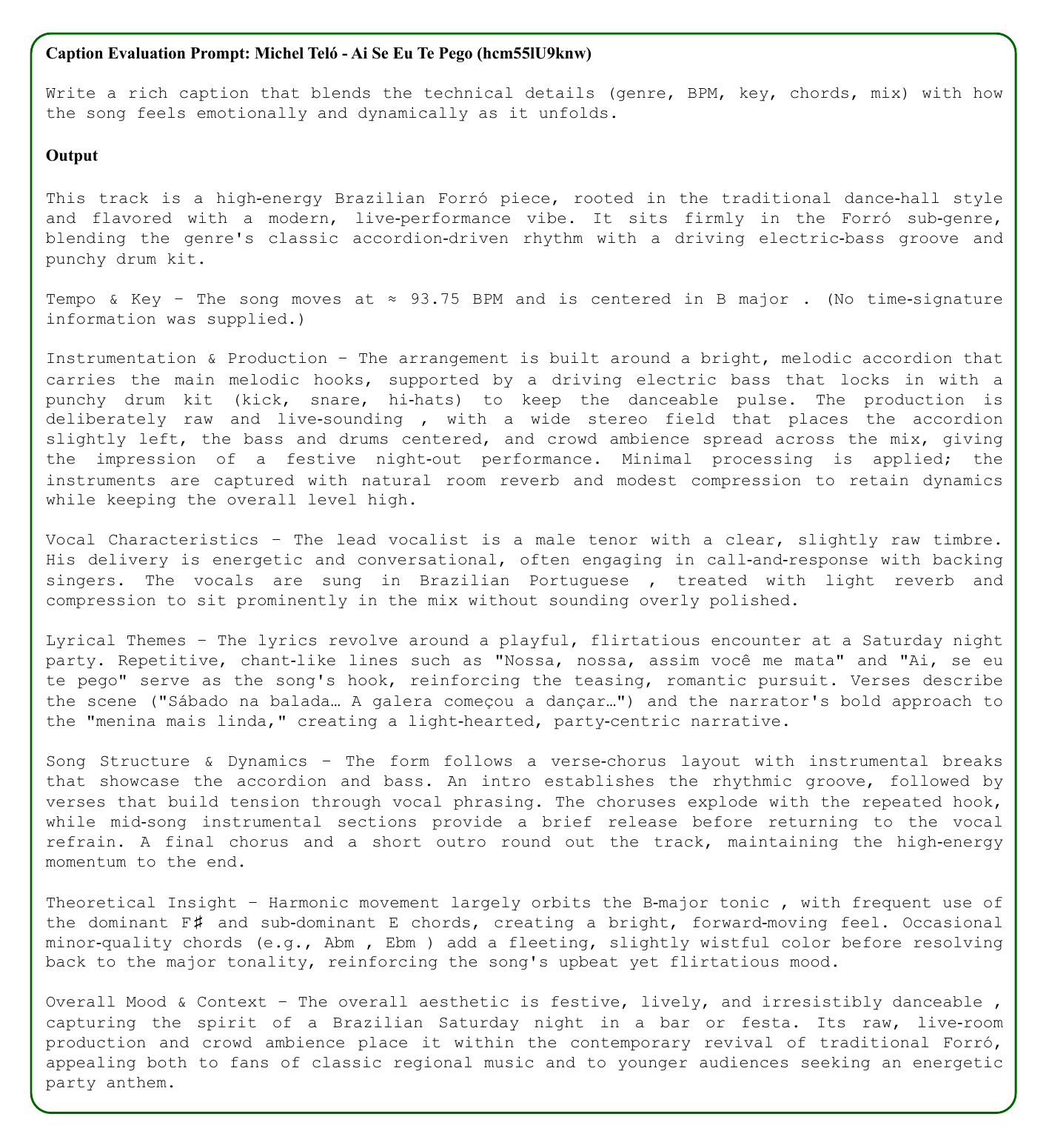}
    \caption{Caption generated by Music Flamingo on a modern Brazilian song.}
    \label{fig.flamingo_brazilian}
    \vspace{-2mm}
\end{figure}

\begin{figure}[!ht]
    \centering
    \includegraphics[width=\linewidth]{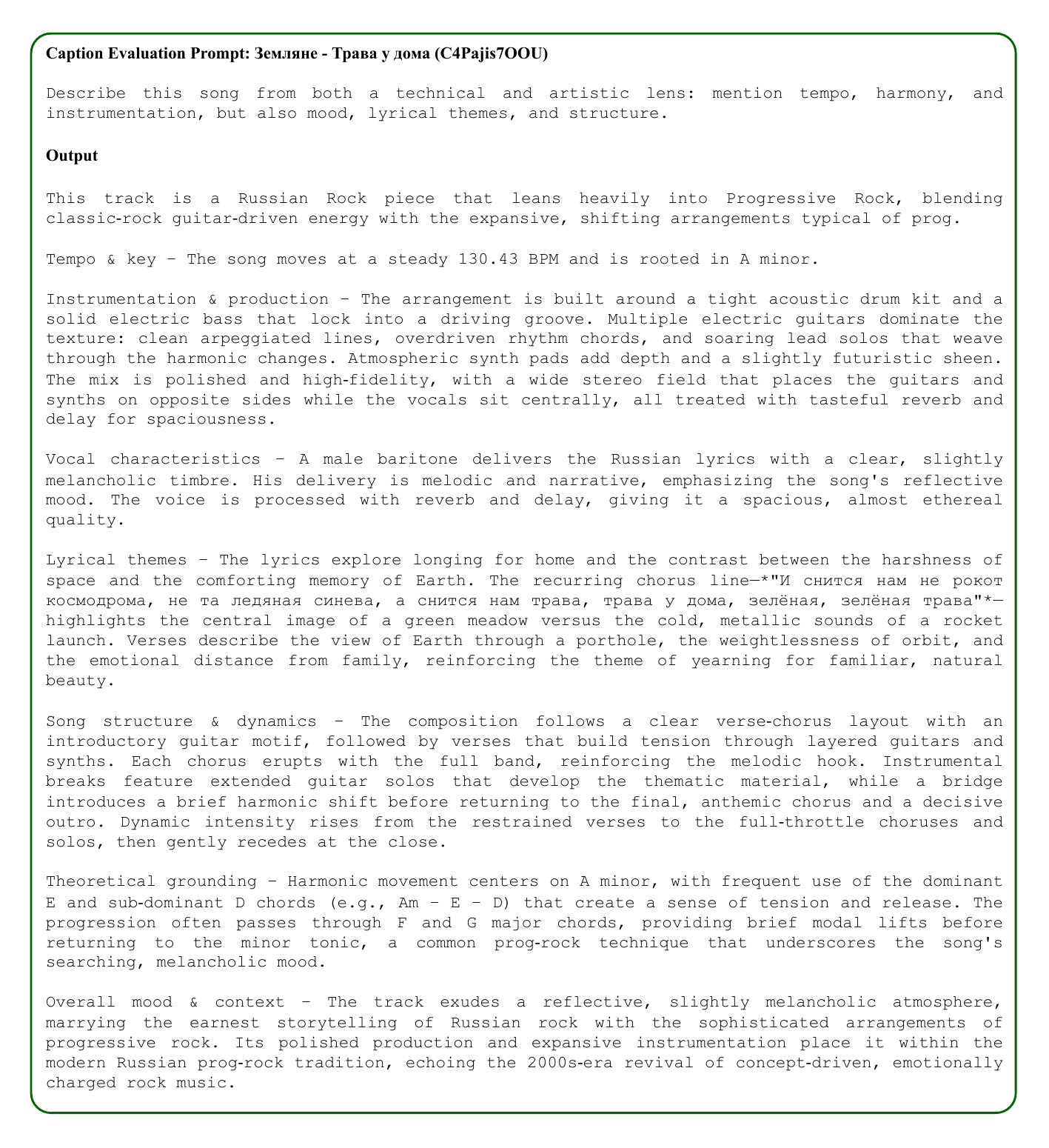}
    \caption{Caption generated by Music Flamingo on a well known Russian song.}
    \label{fig.flamingo_russian}
    \vspace{-2mm}
\end{figure}

Furthermore, we compare the strengths of Music Flamingo on five commercially released songs spanning cultures, languages and styles: Niuver \emph{Enamorados} (Spanish, Latin ballad) (\cref{fig.flamingo_latin}), Annika Wells \emph{Jim \& Pam} (English, indie/acoustic pop)(\cref{fig.flamingo_american}), Louane \emph{La fille} (French, piano-led pop)(\cref{fig.flamingo_french}), Michel Telo \emph{Ai Se Eu Te Pego} (Portuguese, Brazilian sertanejo)(\cref{fig.flamingo_brazilian}), and Zemlyane \emph{Trava u doma} (Russian, Soviet rock)(\cref{fig.flamingo_russian}). Below we present a detailed summary of this comparison. 

\medskip
\noindent\textbf{General technical characteristics (tempo, time signature, key).}
MF consistently produced numeric tempos and keys that matched canonical analyses or widely observed half/double-time readings, and it explicitly handled relative-minor vs.\ metadata-major ambiguities (e.g., \emph{Enamorados}: metadata in C major while harmony centers on A minor). GPT-4o and Gemini often described tempo qualitatively or gave numeric ranges but omitted keys; when numeric BPMs were provided, both models occasionally drifted toward club-tempo values that better reflect remixes than the canonical singles (e.g., \emph{Ai Se Eu Te Pego}: 128--140\,BPM claimed vs.\ $\sim$96\,BPM on the hit version). Qwen3 frequently omitted numerics altogether. Time signature was rarely stated by any model; where implied, 4/4 matched all five tracks.

\emph{Illustrative cases.}
(i) \emph{Jim \& Pam}: MF reported the double-time $\sim$158\,BPM and the correct key (D), aligning with a felt pulse at $\sim$79\,BPM; GPT-4o gave the correct qualitative tempo band but no key; Gemini mis-estimated to 120--125\,BPM and mis-keyed E.
(ii) \emph{La fille}: MF aligned with $\sim$128\,BPM in C; Gemini and GPT-4o underestimated (90--100\,BPM) and/or mis-keyed.
(iii) \emph{Trava u doma}: MF matched A minor and $\sim$130\,BPM; GPT-4o underestimated to $\sim$100--110\,BPM.

\medskip
\noindent\textbf{Genre.}
All models could identify the broad style family. \emph{Gemini} held a \emph{narrow} edge on matching canonical catalog labels and regional taxonomy (e.g., \emph{Ai Se Eu Te Pego}: \emph{sertanejo universit\'{a}rio} with dance-pop trappings). MF was directionally correct across the set and, in two cases, selected closely related tags when salient timbres or arrangement scale were misleading (\emph{Ai Se Eu Te Pego}: forr\'{o} inferred from accordion timbre; \emph{Trava u doma}: prog-leaning language for a synth/space-colored Soviet pop/rock record). Importantly, these adjacent picks did not derail MF’s downstream harmony/structure reasoning and are straightforward to normalize to catalog labels. \emph{GPT-4o} typically described the stylistic \emph{feel} accurately (e.g., dance-pop with Brazilian flair; piano-led ballad) but often stopped short of naming the canonical label. \emph{Qwen3} alternated between sensible tags (French indie/pop ballad) and broad era styles (``80s arena/soft rock''), and twice misframed vocal songs as instrumental (see below), which contaminated the subsequent genre claim.

\emph{Takeaway.} Canonical label accuracy: \textbf{Gemini} $\gtrsim$ \textbf{MF} $\approx$ \textbf{GPT-4o} $>$ \textbf{Qwen3}. Gemini’s advantage is mostly in verbatim catalog taxonomy; MF’s labels are correct at the family level and, when adjacent, remain musicologically consistent with its superior harmonic/structural analysis.

\medskip
\noindent\textbf{Emotional content and lyrics.}
MF, Gemini, and GPT-4o gave coherent, text-grounded readings of mood and themes across all songs (e.g., \emph{Enamorados}: memory, time, and fading love; \emph{La fille}: identity and self-doubt; \emph{Trava u doma}: homesick cosmonaut narrative). Where lyrics were quoted or paraphrased, all three remained faithful to content and tone. Qwen3 produced reasonable affect reads when it acknowledged lyrics, but twice declared a vocal track ``instrumental'' (\emph{Enamorados}, \emph{Trava u doma}), leading to incorrect conclusions about narrative and emotion.

\emph{Observation.} When models inferred emotion strictly from sonics without anchoring in lyric text, nuance decreased and culture-specific references were missed (e.g., \emph{Trava u doma} as an iconic space-age anthem; \emph{Ai Se Eu Te Pego} as a global sertanejo earworm driven by chant-like hooks).

\medskip
\noindent\textbf{Complex technical characteristics (chord progressions/voicings, song structure, production).}
MF generally provided the deepest harmonic/structural content (naming cadential behavior, relative-minor centers, verse/chorus dynamics), and its structural reads were consistently plausible across all five songs. Its main failure mode was \emph{over-specification}: occasionally asserting colorful altered/extended chords or percussion layers not supported by public charts or by the stems one would expect (\emph{Jim \& Pam}: introduced drum-machine and synth-bass in an otherwise hand-clap/snaps, acoustic texture; \emph{La fille}: added brushed kit and altered dominants to a piano-centric, drum-light mix). GPT-4o’s arrangements and sectioning were reliably correct (intro/verse/chorus/bridge placement, dynamic swells), with conservative but accurate production notes; it rarely named specific harmonic content, which limited precision but avoided hallucination. Gemini’s arrangement commentary was serviceable and sometimes quite apt on famous tracks (accordion/synth hook in \emph{Ai Se Eu Te Pego}), but often remained generic and light on concrete harmony. Qwen3’s technical layer was the sparsest and suffered when the top-level premise was wrong (``instrumental''), cascading into inapplicable structure/production claims.

\emph{Failure-mode coupling.} We repeatedly observed that \textbf{genre misclassification leads to production hallucinations}. For instance, mapping \emph{Ai Se Eu Te Pego} to \emph{forr\'{o}} primed mentions of forr\'{o}-typical percussion, and reading \emph{La fille} as an indie/pop ballad with a ``soft electronic beat'' invited non-existent drum programming. Conversely, when models named the \emph{canonical} genre, instrumentation and mix notes tended to be accurate (Gemini on \emph{Ai Se Eu Te Pego}; MF and GPT-4o on the piano+voice core of \emph{La fille}).

\medskip
\noindent\textbf{Model-specific observations (holistic).}
\begin{itemize}
  \item \textbf{Music Flamingo.} Strongest on measurable facts (tempo/key) and the only model to consistently reconcile metadata major vs.\ relative-minor centers (\emph{Enamorados}: C vs.\ A minor). Best overall at section-level reasoning and harmonic intent, but occasionally over-decorates with color chords or speculative percussion.
  \item \textbf{Gemini 2.5 Pro.} Best at canonical genre taxonomy and culturally grounded framing (e.g., \emph{sertanejo universit\'{a}rio}). Tends to under-specify harmony and sometimes over-estimate tempo; instrumentation is usually correct when genre is correct.
  \item \textbf{GPT-4o.} Most dependable for arrangement, dynamics, and production prose; excellent at emotion/lyric grounding. Hesitant on numerics (tempo/key) and sometimes avoids naming the canonical genre even when its prose implies it.
  \item \textbf{Qwen3-Omni.} Capable of coherent mood and arrangement reads when anchored, but inconsistent. The two ``instrumental'' misclassifications (on clearly vocal songs) demonstrate a brittle failure that propagates into wrong genre/production claims.
\end{itemize}

\medskip
\noindent Overall, the pattern is clear: \emph{precision on the measurable (tempo/key) + correct canonical genre naming} is the foundation for faithful structure/production descriptions and culturally aware, lyric-grounded emotion reads. Among the models studied, \textbf{Music Flamingo} delivers the most reliable technical grounding and music-theoretic reasoning; \textbf{Gemini} adds the most accurate taxonomy; and \textbf{GPT-4o} supplies consistently correct structural/production narratives.
% End Lasha's version

\newpage
\newpage
\newpage

\newpage
\newpage

\section{Linear probing experiments with audio encoders}
\label{sec.linear_probing}
In order to better understand the role of the audio encoder in music tasks, we performed linear probing experiments with three audio encoders: the audio encoder from Qwen2Audio~\citep{chu2024qwenaudio2} and Audio Flamingo 3~\citep{goel2025audioflamingo3advancing} -- the encoder on which Music Flamingo is based, which are both based on the Whisper architecture, and the MERT encoder~\citep{yizhi2023mert}, which has an architecture with a bias towards understanding music due to its use of a constant-Q transform representation in the input. We chose two tasks from the MARBLE benchmark \citep{yuan2023marble}: the key classification task using the GS dataset \citep{knees2015gs}, and the genre classification task using the GTZAN dataset \citep{tzanetakis2002gtzan}. All linear probing models were trained using the average of all the frames from the audio representation in question (Qwen2Whisper, AFWhisper, or MERT) with a single linear layer. Models were trained until no improvement in validation set accuracy was observed after 5 epochs, and validation set accuracy was used as well to choose the best checkpoint. Results can be found in \Cref{tab:linear_probing}.

\begin{table}[h]
\centering
\caption{\small Performance comparison on key and genre classification with different audio encoders. }
\vspace{2mm}
\resizebox{0.5\linewidth}{!}{
\begin{tabular}{lcc}
\toprule
\textbf{Model} & \textbf{GS (key cls.)↑} & \textbf{GTZAN (genre cls.)↑} \\
\midrule
Qwen2Audio & 34.10 & 89.99 \\
AFWhisper & 40.56 & \textbf{91.37} \\
MERT & \textbf{56.12} & 78.96 \\
\bottomrule
\end{tabular}
}
\label{tab:linear_probing}
\end{table}

We observe that both models based on Whisper, which were trained specifically for captioning, have high accuracy for genre classification but comparatively low accuracy for the key classification task. We hypothesize this is because when using captioning targets, the likelihood of having the genre for a given song mentioned in the caption is much higher than having the key the song is in. MERT, on the other hand, has lower accuracy for the higher level task but higher accuracy for the lower level task. MERT is trained to generate a representation that can be used to reconstruct the audio through a VAE "acoustic teacher" model, and also to match the coefficients of a CQT transform (the "music model"), therefore being a more lower level representation than text-based captions.

Given the correlation between this gap in accuracy between lower and higher level tasks and the general observations for the outputs of Music Flamingo, in the future we intend to try an audio encoder that preserves more of the lower level information in audio to enable future versions of the model to perform better in understanding those features from music signals.
% \pagebreak
\subsection{Examples of MF-Skills and MF-Think}
\label{sec.mf_examples}
\label{subsec:examples_mfthink}
\begin{figure}[!h]
    \centering
    \includegraphics[width=\linewidth]{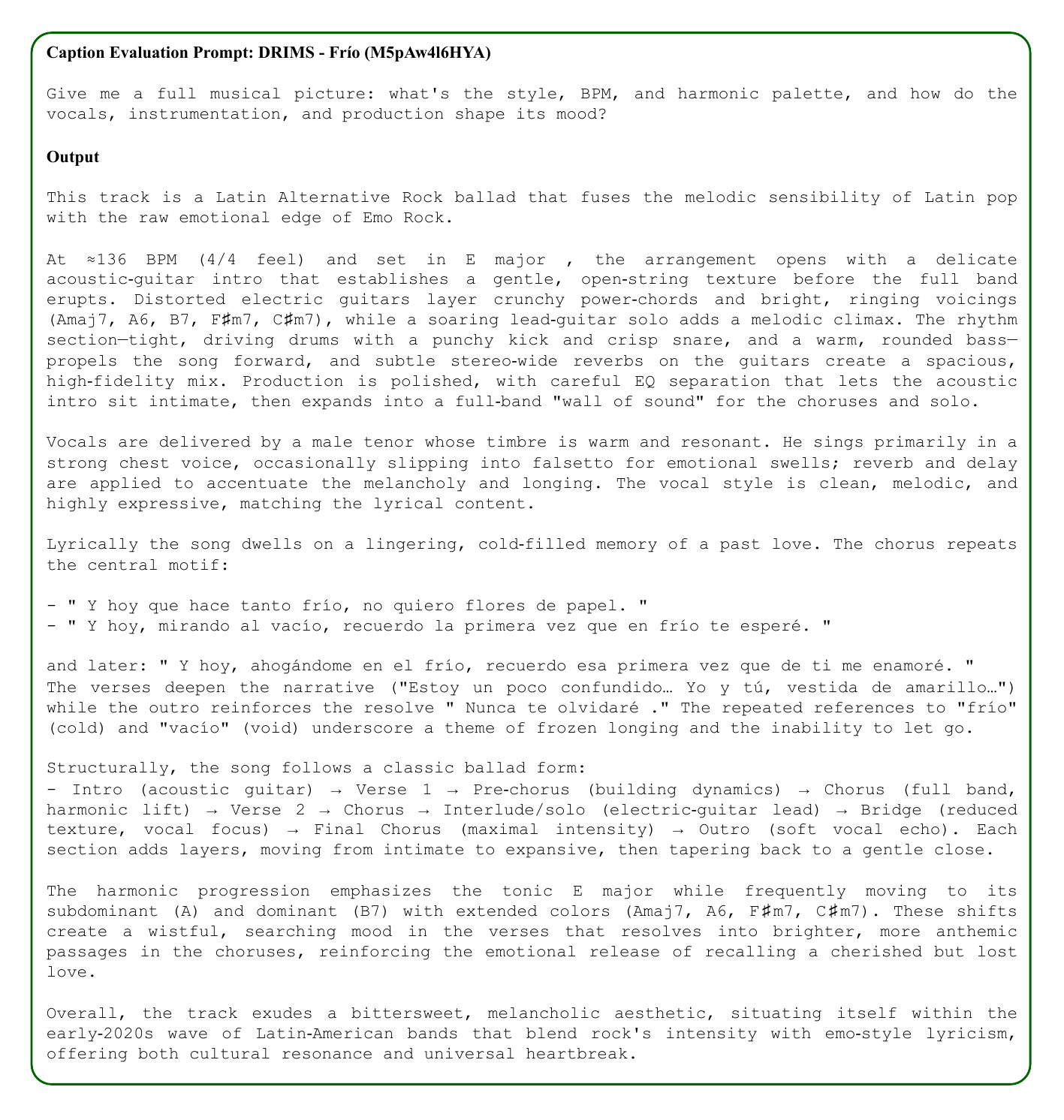}
    \caption{Example of a ground-truth caption on a Latin song in MF-Skills.}
    \label{fig.mf-skills_1}
    \vspace{-2mm}
\end{figure}

\begin{figure}[!ht]
    \centering
    \includegraphics[width=\linewidth]{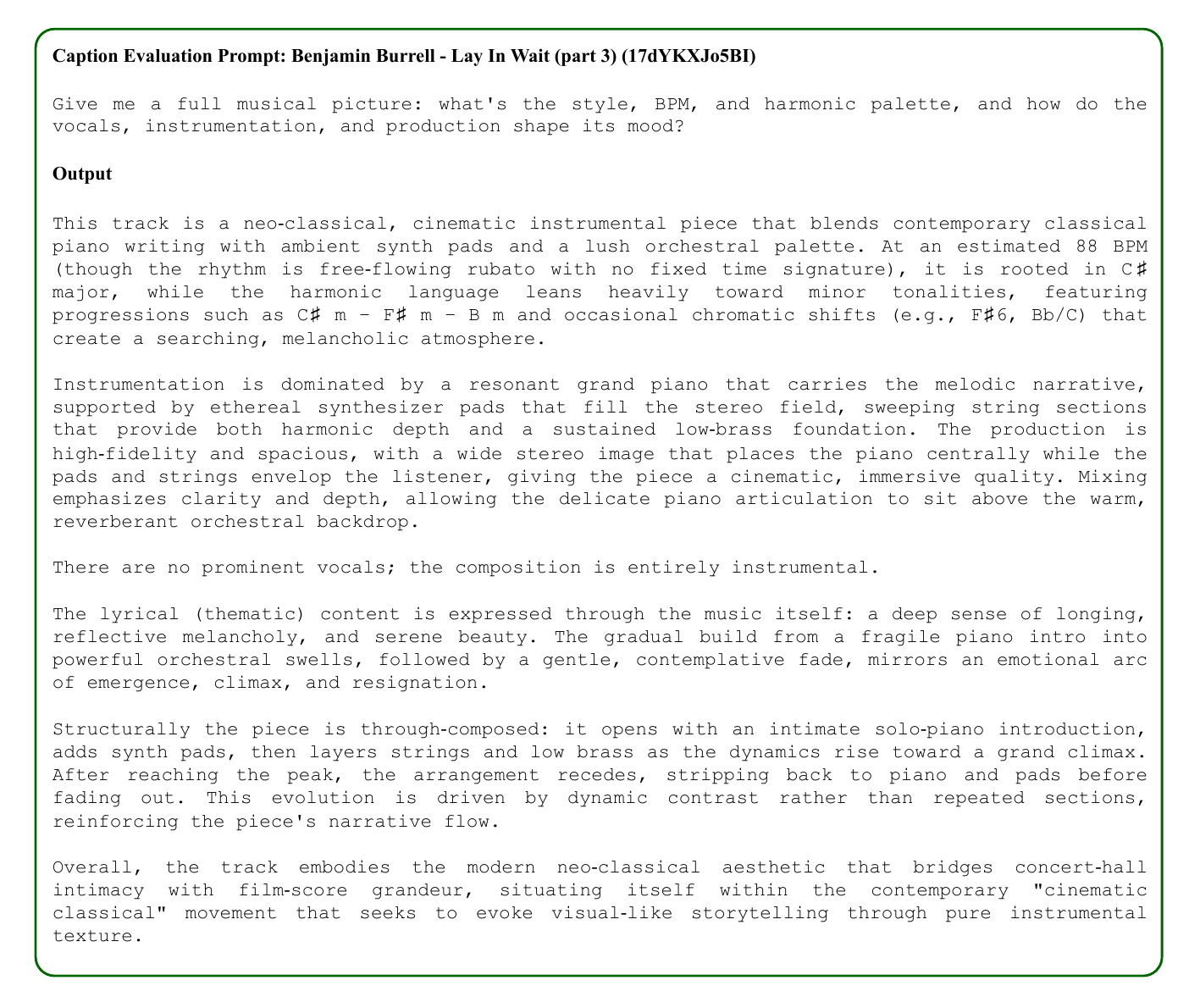}
    \caption{Example of a ground-truth caption on a classical instrumental piece in MF-Skills.}
    \label{fig.mf-skills_2}
    \vspace{-2mm}
\end{figure}

\begin{figure}[!ht]
    \centering
    \includegraphics[width=\linewidth]{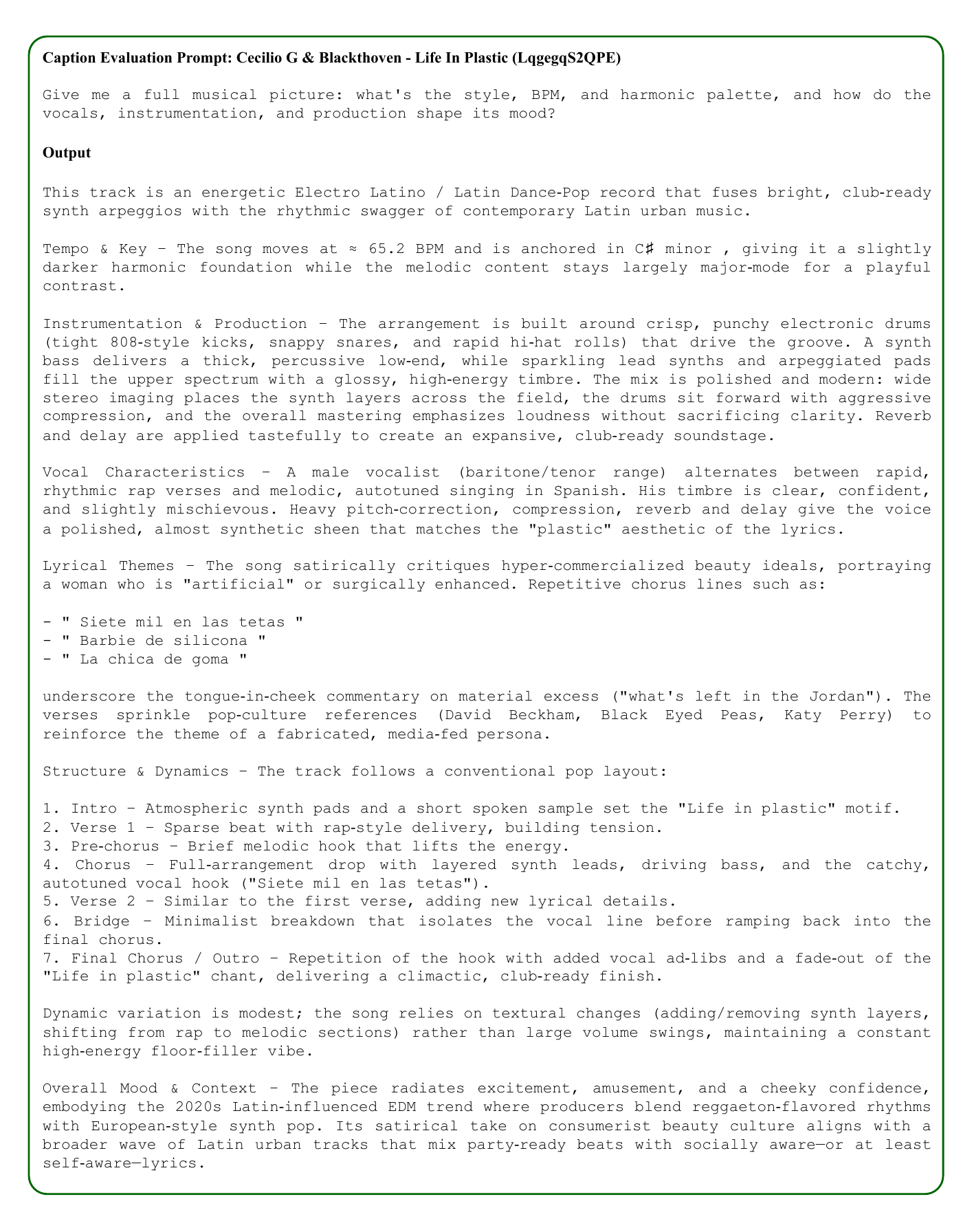}
    \caption{Example of a ground-truth caption on a Latin song in MF-Skills.}
    \label{fig.mf-skills_3}
    \vspace{-2mm}
\end{figure}

\begin{figure}[!ht]
    \centering
    \includegraphics[width=\linewidth]{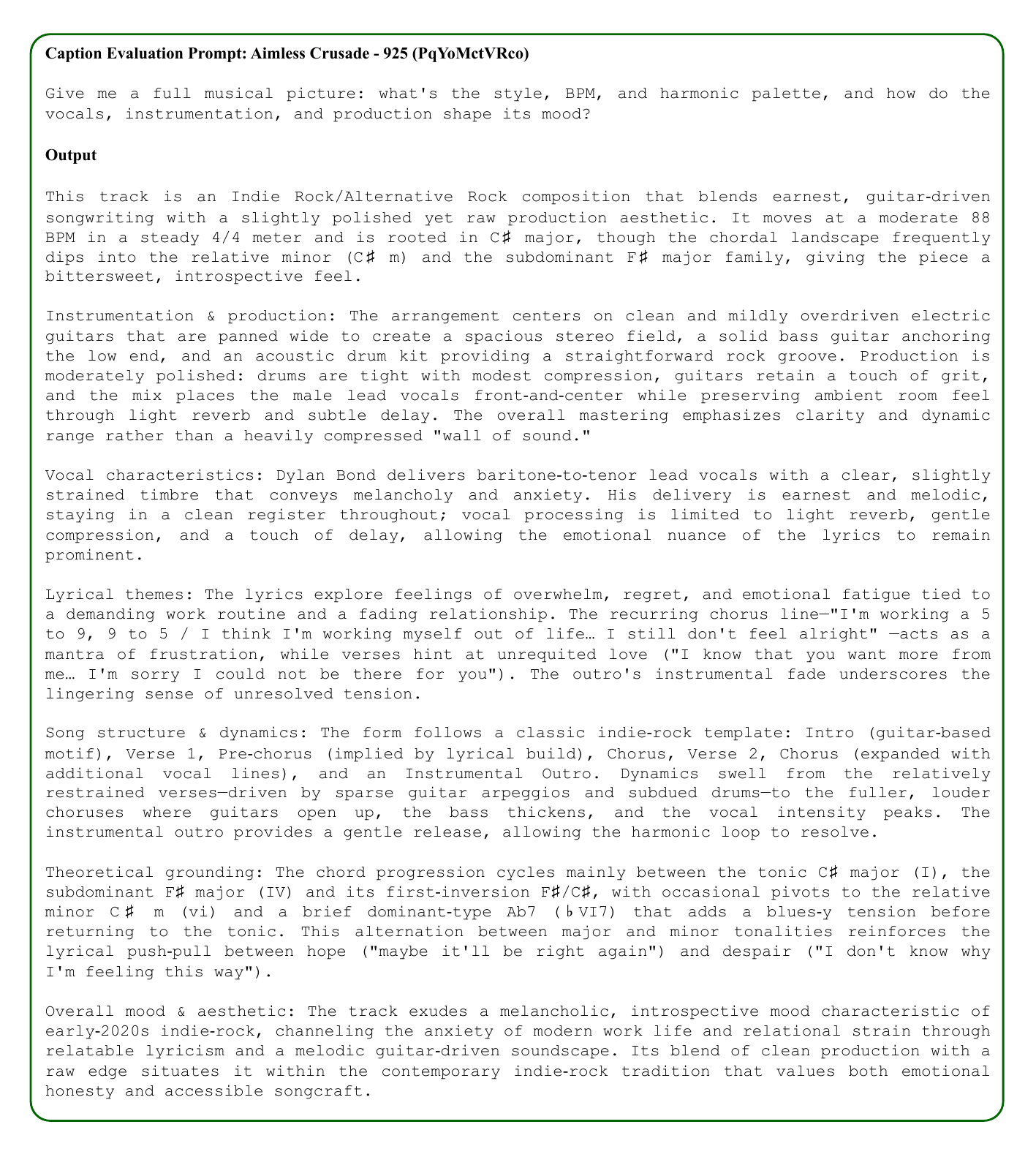}
    \caption{Example of a ground-truth caption on an American song in MF-Skills.}
    \label{fig.mf-skills_4}
    \vspace{-2mm}
\end{figure}

\begin{figure}[!ht]
    \centering
    \includegraphics[width=\linewidth]{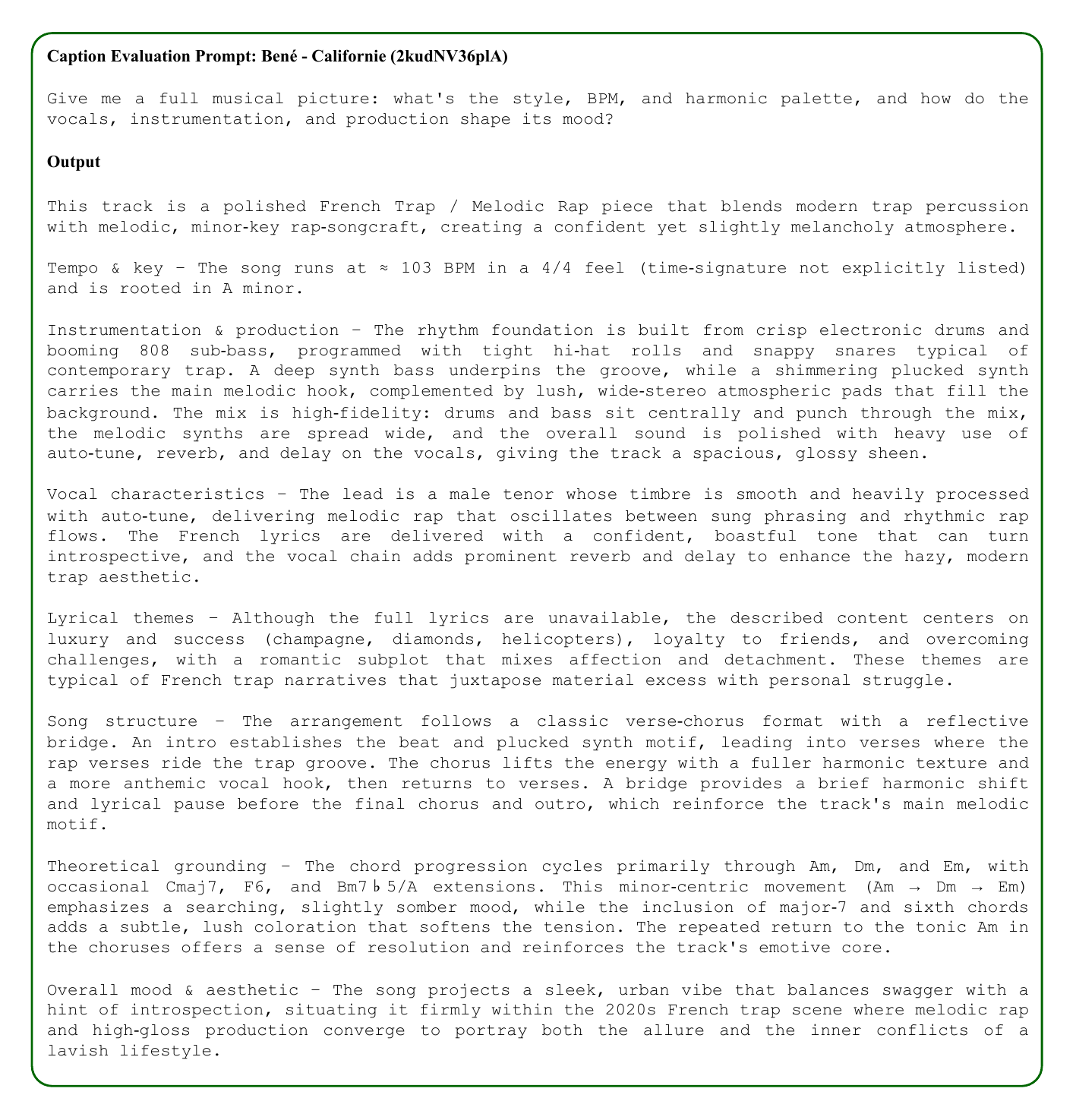}
    \caption{Example of a ground-truth caption on a French song in MF-Skills.}
    \label{fig.mf-skills_5}
    \vspace{-2mm}
\end{figure}

\begin{figure}[!ht]
    \centering
    \includegraphics[width=\linewidth]{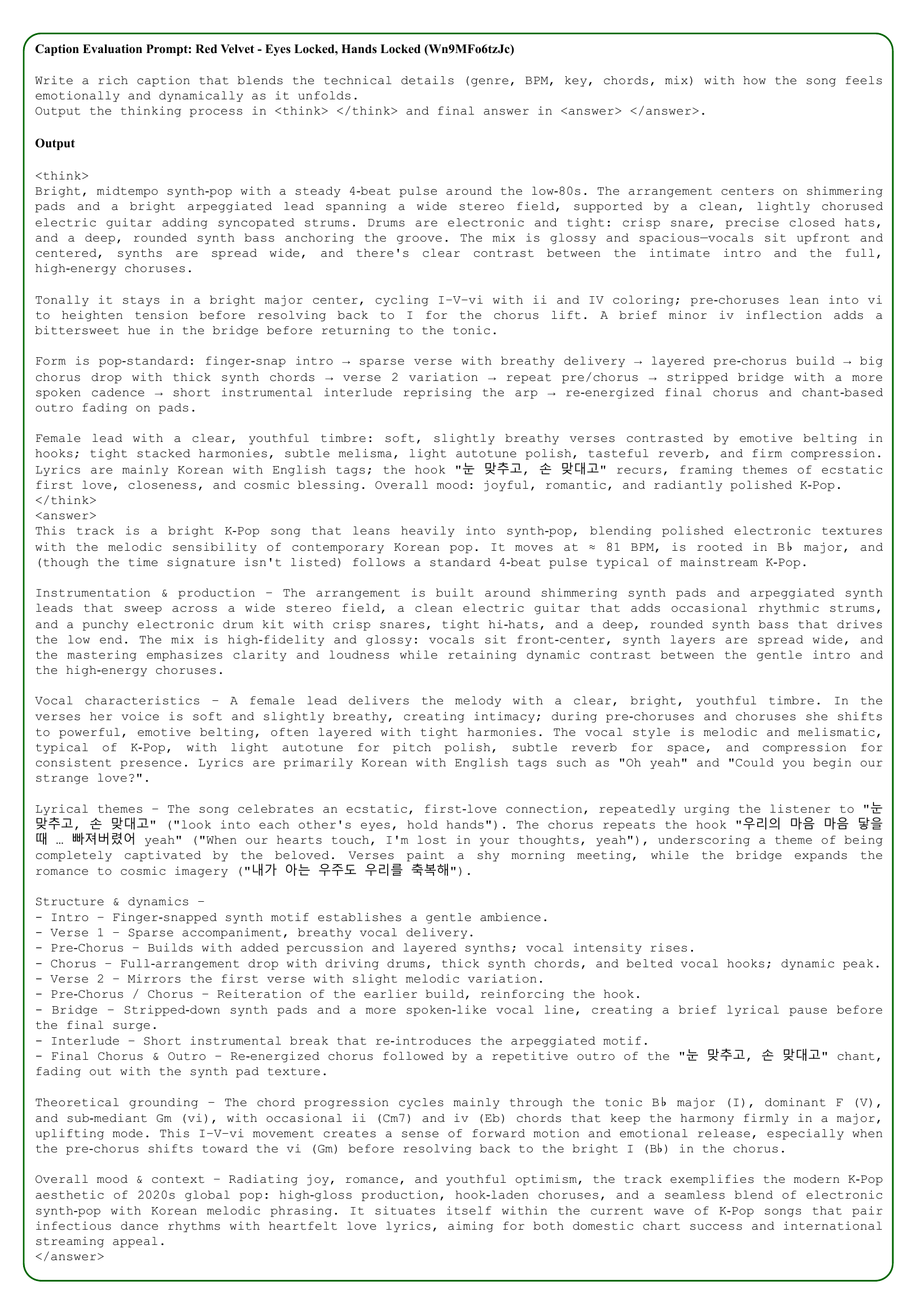}
    \caption{Example of a ground-truth caption with thinking traces on a Korean song in MF-Think.}
    \label{fig.mf-think-captions_1}
    \vspace{-2mm}
\end{figure}

\begin{figure}[!ht]
    \centering
    \includegraphics[width=\linewidth]{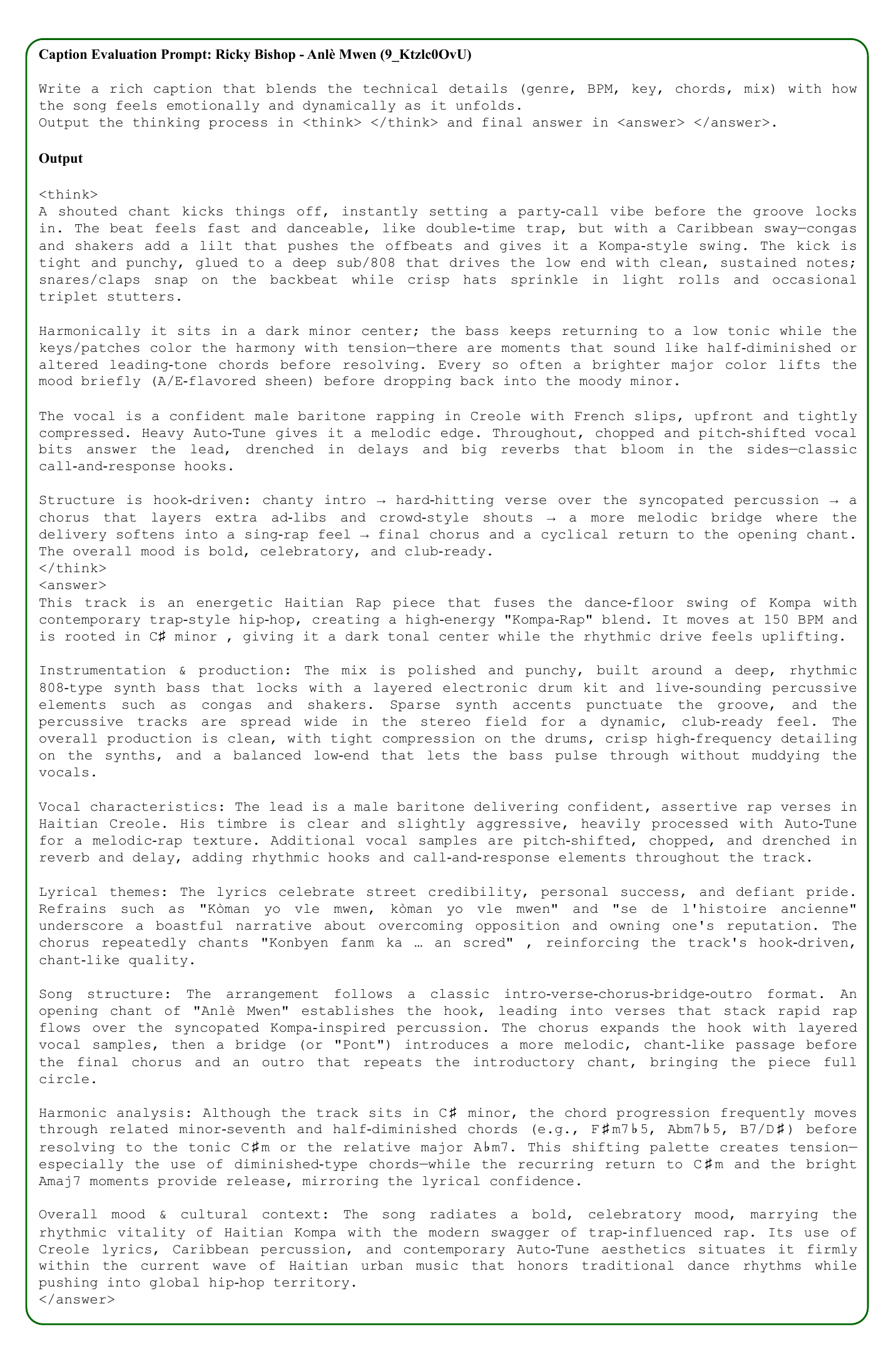}
    \caption{Example of a ground-truth caption with thinking traces on a Haitian song in MF-Think.}
    \label{fig.mf-think-captions_2}
    \vspace{-2mm}
\end{figure}

\begin{figure}[!t]
    \centering
    \includegraphics[width=\linewidth]{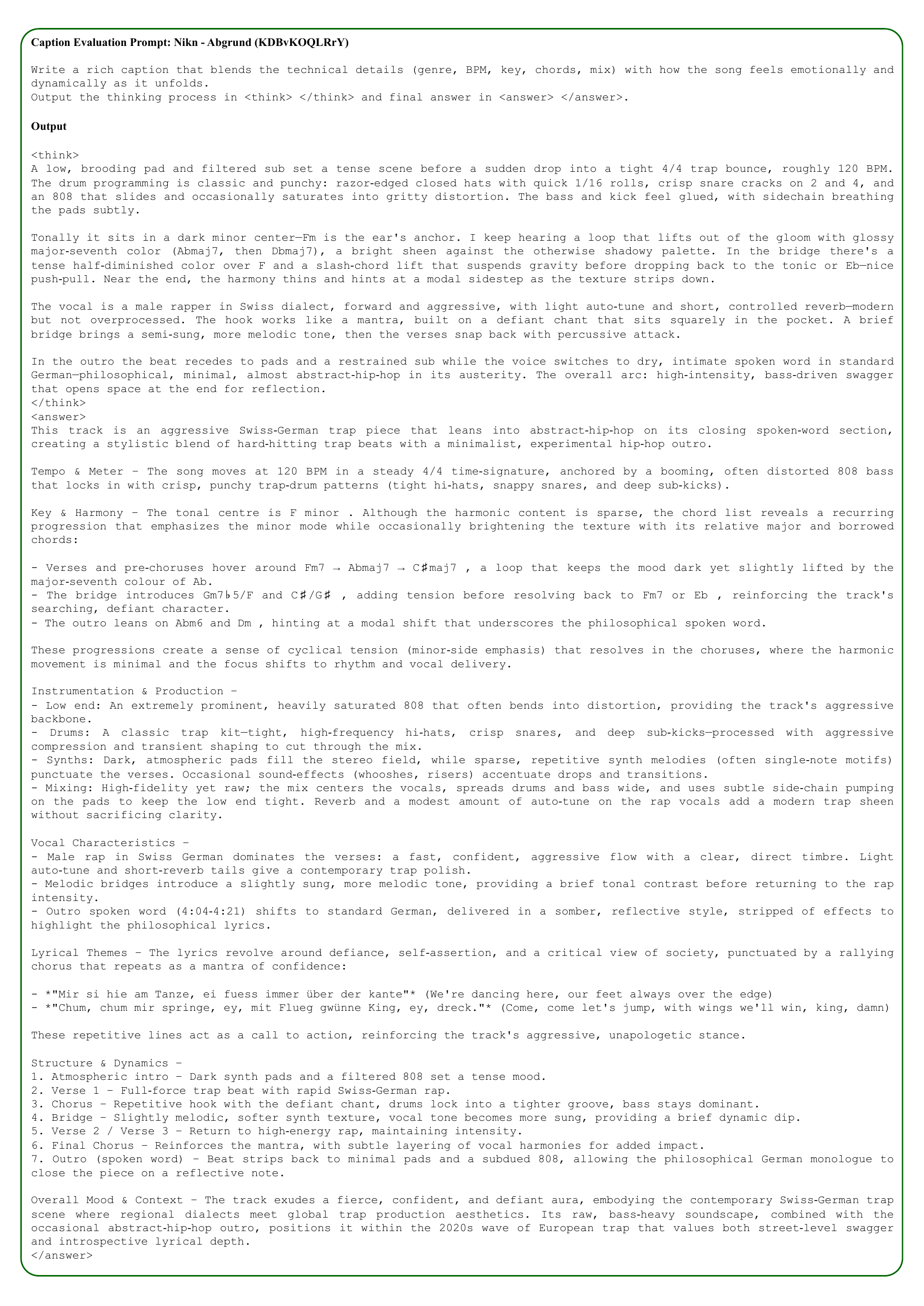}
    \caption{Example of a ground-truth caption with thinking traces on a European song in MF-Think.}
    \label{fig.mf-think-captions_3}
    \vspace{-2mm}
\end{figure}

\begin{figure}[!ht]
    \centering
    \includegraphics[width=\linewidth]{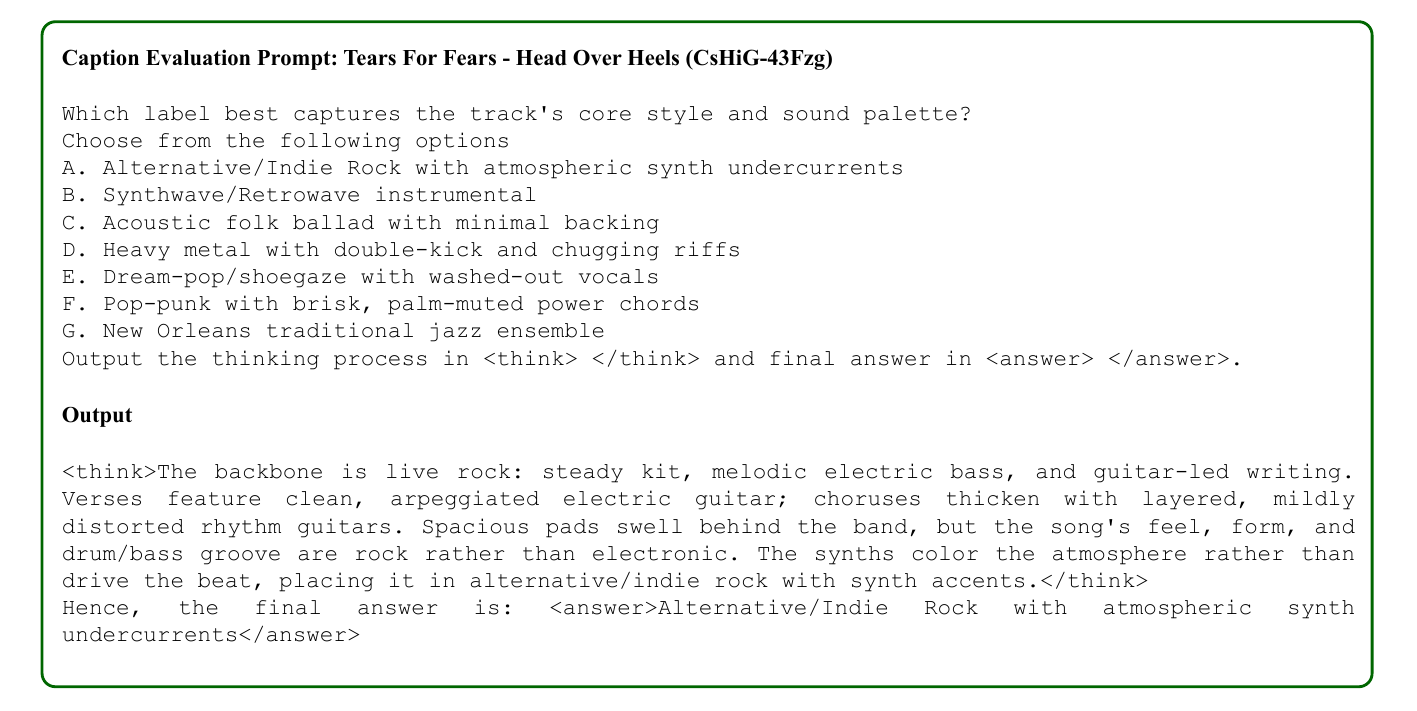}
    \caption{Example of QA pairs with thinking traces in MF-Think.}
    \label{fig.mf-think-qa_1}
    \vspace{-2mm}
\end{figure}

\begin{figure}[!ht]
    \centering
    \includegraphics[width=\linewidth]{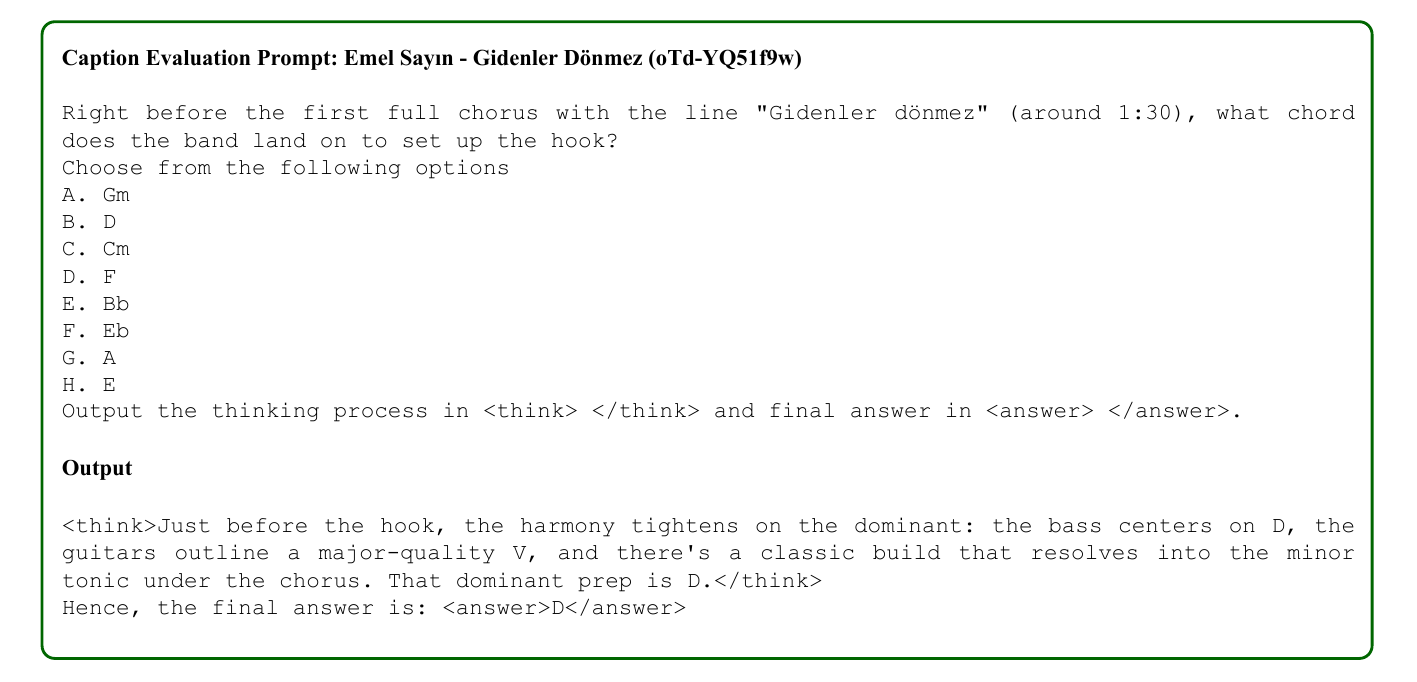}
    \caption{Example of QA pairs with thinking traces in MF-Think.}    \label{fig.mf-think-qa_2}
    \vspace{-2mm}
\end{figure}

\begin{figure}[!ht]
    \centering
    \includegraphics[width=\linewidth]{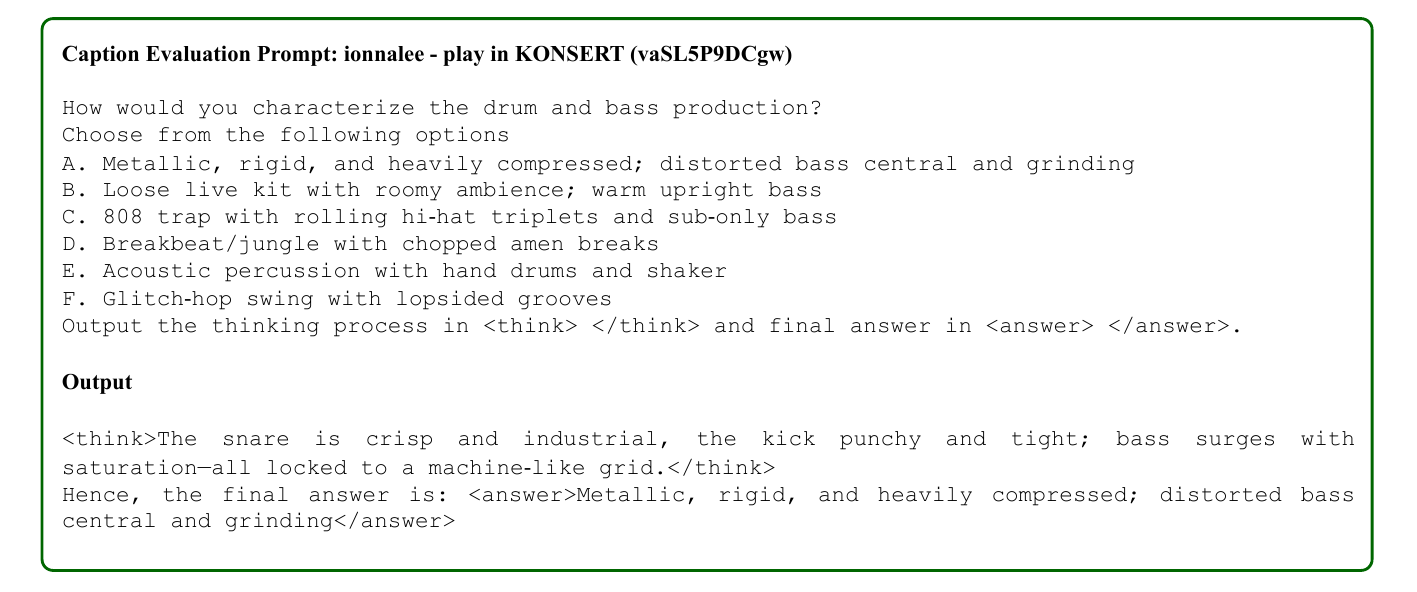}
    \caption{Example of QA pairs with thinking traces in MF-Think.}
    \label{fig.mf-think-qa_3}
    \vspace{-2mm}
\end{figure}

\emph{MF-Skills} contains long, multi-aspect targets (captions and QA) \emph{without} reasoning traces. \Cref{fig.mf-skills_1},~\Cref{fig.mf-skills_2},~\Cref{fig.mf-skills_3},~\Cref{fig.mf-skills_4},~\Cref{fig.mf-skills_5} shows examples of captions in the MF-Skills dataset. \emph{MF-Think} augments a challenging subset with compact chain-of-thought traces (\texttt{<think>}\dots\texttt{</think>}) and final answers (\texttt{<answer>}\dots\texttt{</answer>}). \Cref{fig.mf-think-captions_1},~\Cref{fig.mf-think-captions_2},~\Cref{fig.mf-think-captions_3},~\Cref{fig.mf-think-qa_1},~\Cref{fig.mf-think-qa_2} and ~\Cref{fig.mf-think-qa_3} shows examples of captions and QA pairs in the MF-Think dataset.
% \textbf{How to read the examples.} In Fig.~\ref{fig:mfskills_mfthink_examples_onecol}, each example is printed in a single column to fit variable-length text. The MF-Skills target (no CoT) is followed by the MF-Think reasoning snippet and final answer. Full-length examples appear in Appendix~B.1.

\newpage
\newpage
\newpage
\newpage
\newpage
\newpage
\newpage
\newpage
\newpage
\newpage

\pagebreak
\section{Prompts for MF-Skills and MF-Think}
\label{sec.prompts}
\begin{figure}[!t]
    \centering
    \includegraphics[width=\linewidth]{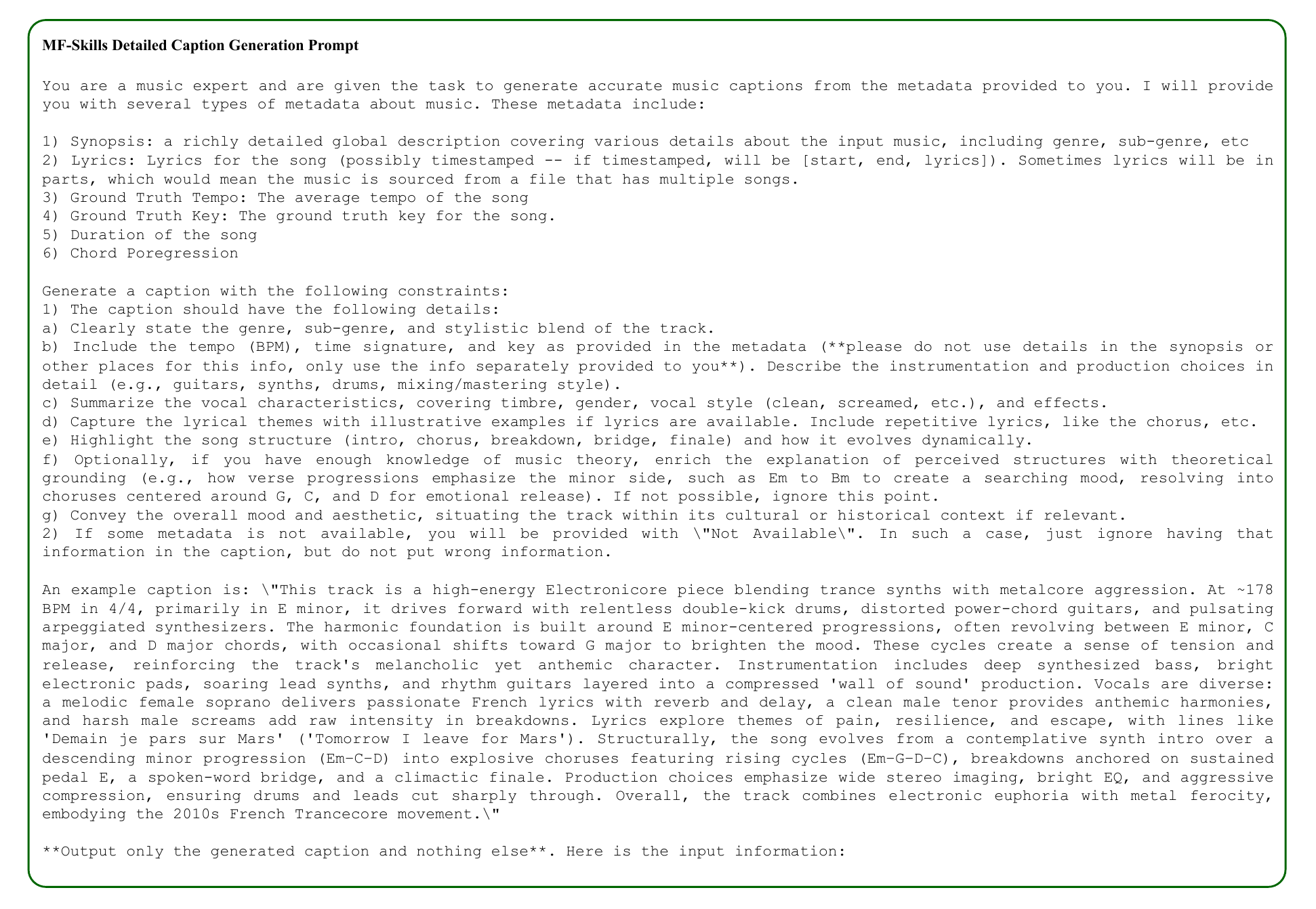}
    \caption{\small Prompt for generating detailed captions for the MF-Skills dataset.}
    \label{fig.prompt5}
    \vspace{-2mm}
\end{figure}

\begin{figure}[!t]
    \centering
    \includegraphics[width=\linewidth]{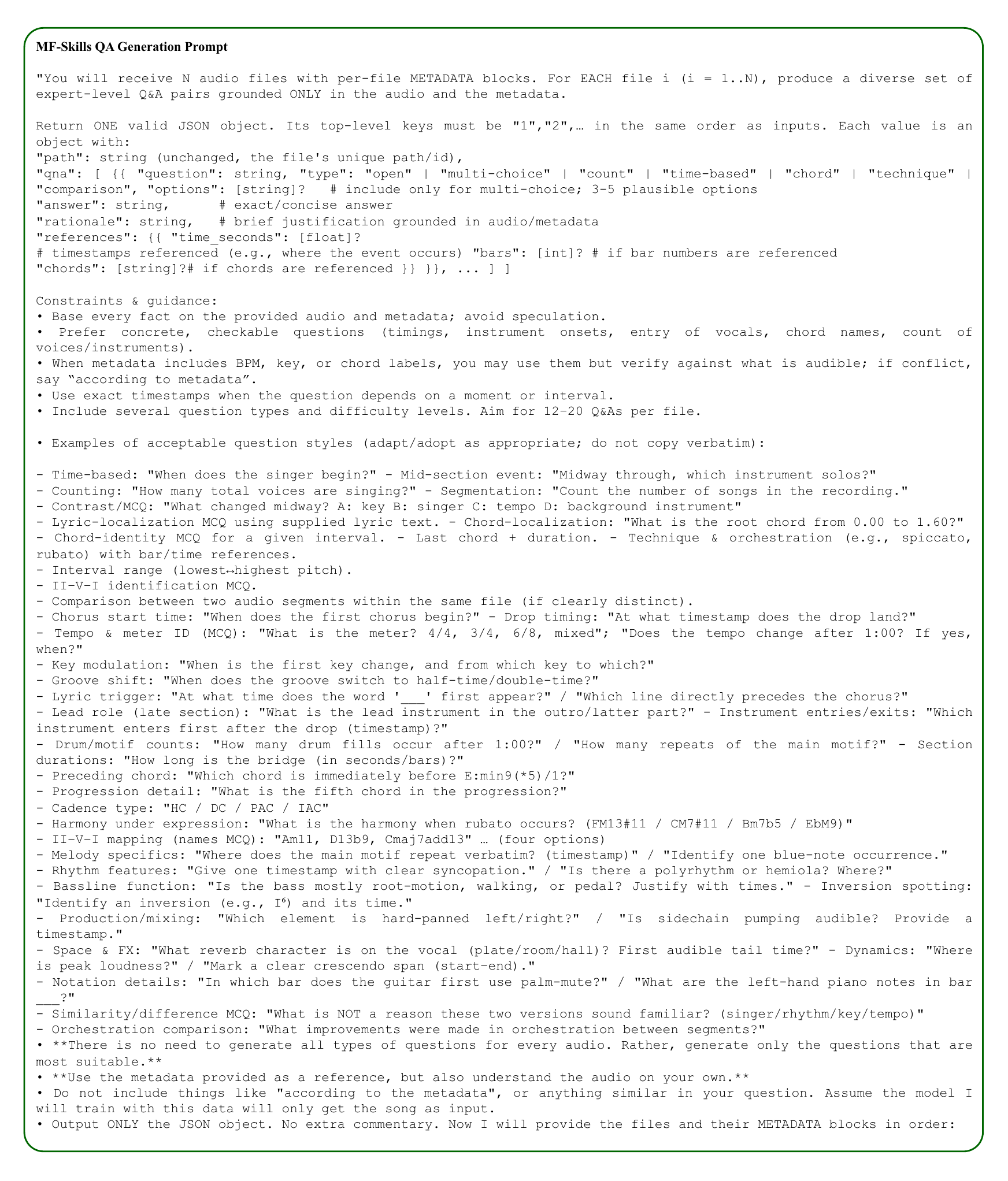}
    \caption{\small Prompt for generating QA pairs for the MF-Skills dataset.}
    \label{fig.prompt1}
    \vspace{-2mm}
\end{figure}

\begin{figure}[!t]
    \centering
    \includegraphics[width=\linewidth]{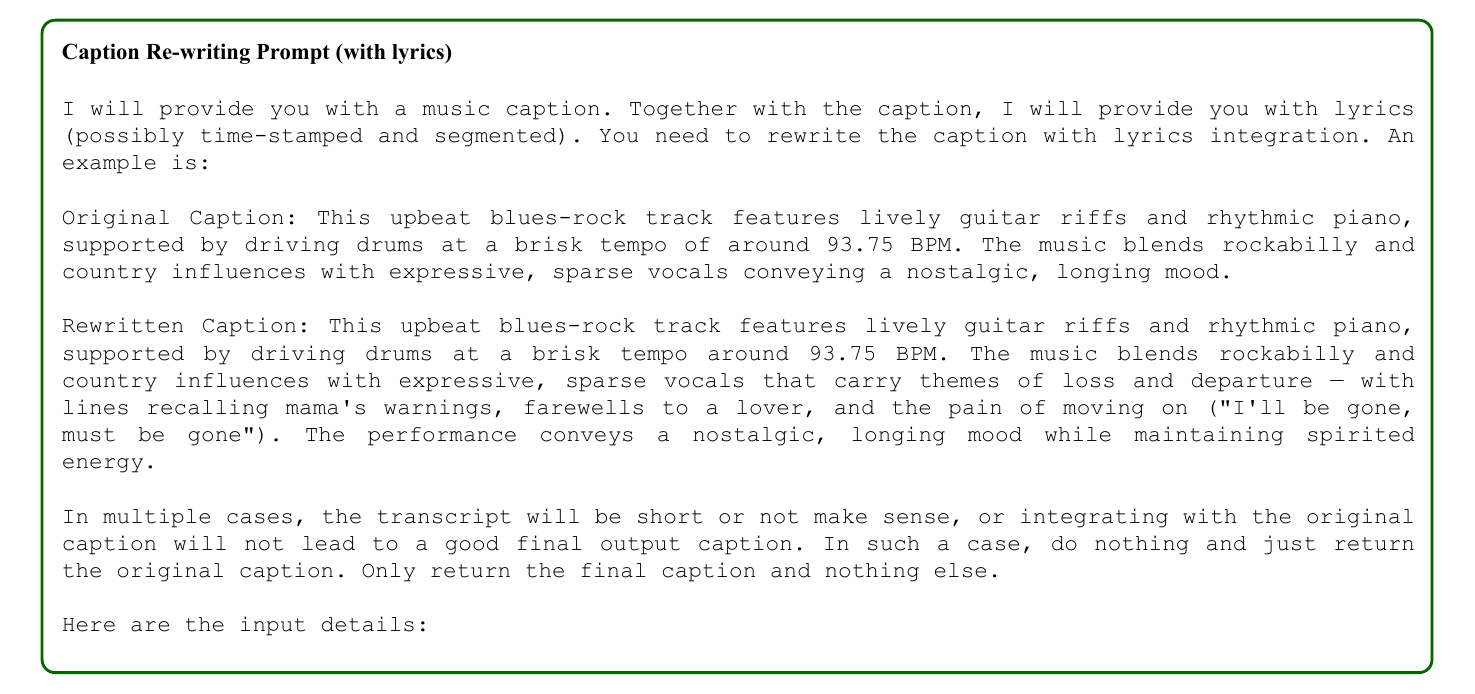}
    \caption{\small Prompt for correcting existing captions with lyrics and metadata on Music4ALL and MSD datasets.}
    \label{fig.prompt2}
    \vspace{-2mm}
\end{figure}

\begin{figure}[!t]
    \centering
    \includegraphics[width=\linewidth]{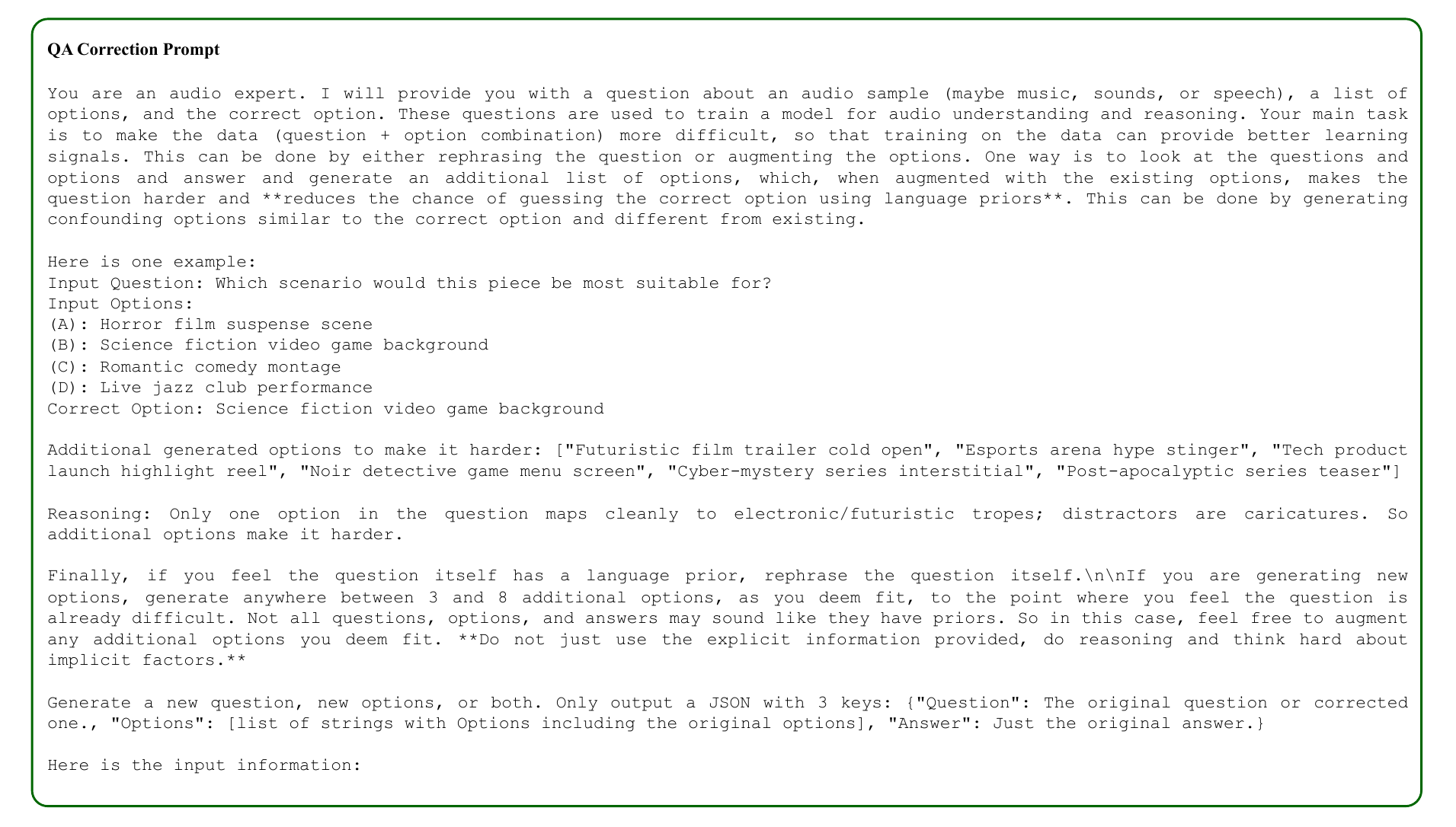}
    \caption{\small Prompt for option augmentation of existing question-answer pairs on the Music4ALL and MSD datasets.}
    \label{fig.prompt4}
    \vspace{-2mm}
\end{figure}

\begin{figure}[!t]
    \centering
    \includegraphics[width=\linewidth]{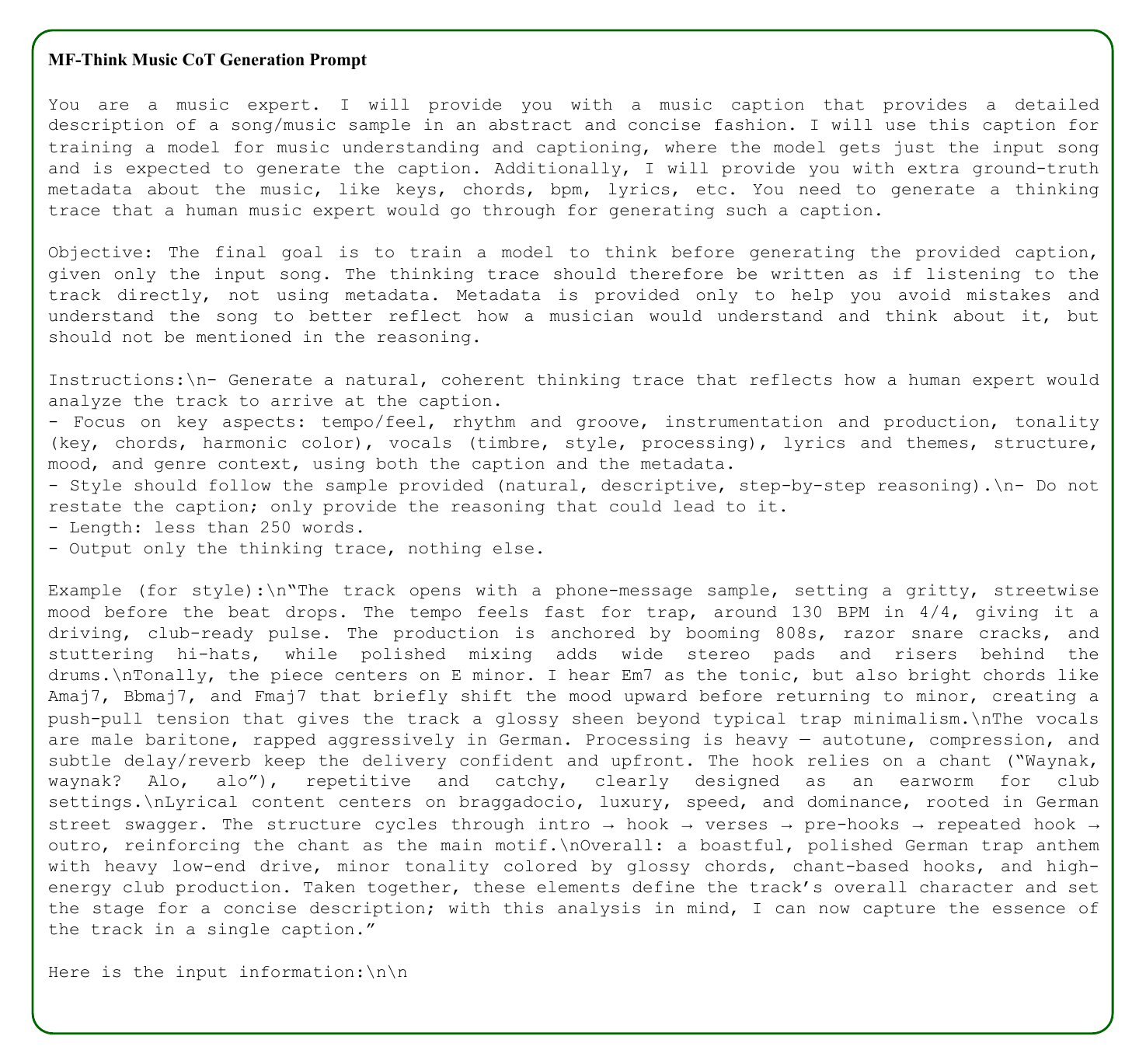}
    \caption{\small Prompt for generating step-by-step reasoning for the detailed captions in the MF-Think dataset.}
    \label{fig.prompt6}
    \vspace{-2mm}
\end{figure}

\begin{figure}[!t]
    \centering
    \includegraphics[width=\linewidth]{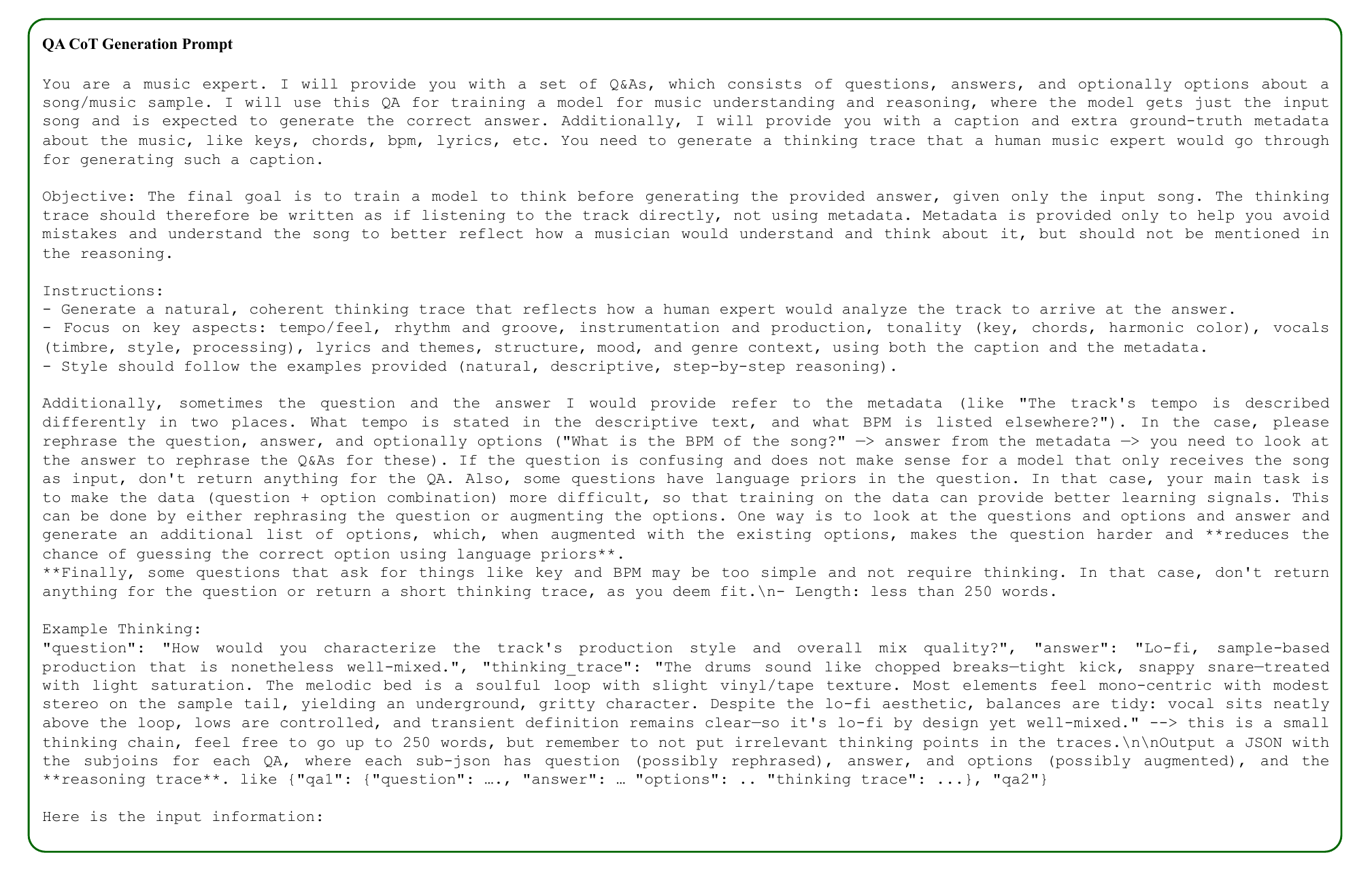}
    \caption{\small Prompt for generating step-by-step reasoning for the question-answer pairs in the MF-Think dataset.}
    \label{fig.prompt3}
    \vspace{-2mm}
\end{figure}

We provide all prompting templates used across our datasets and QA types in Figures~\ref{fig.prompt5},~\ref{fig.prompt1},~\ref{fig.prompt2},~\ref{fig.prompt4},~\ref{fig.prompt6},~\ref{fig.prompt3}.

\newpage

\newpage
\newpage

\newpage

\end{document}